\documentclass[12pt,preprint]{aastex}
\usepackage{lscape}
\newcommand{\mi}{\relax \ifmmode {\mu{\mbox m}}\else $\mu$m\fi}
\newcommand{\hii}{\relax \ifmmode {\mbox H\,{\scshape ii}}\else H\,{\scshape ii}\fi}
\newcommand{\sii}{\relax \ifmmode {\mbox S\,{\scshape ii}}\else S\,{\scshape ii}\fi}
\newcommand{\nii}{\relax \ifmmode {\mbox N\,{\scshape ii}}\else N\,{\scshape ii}\fi}
\newcommand{\oiii}{\relax \ifmmode {\mbox O\,{\scshape iii}}\else O\,{\scshape iii}\fi}
\newcommand{\ha}{\relax \ifmmode {\mbox H}\alpha\else H$\alpha$\fi}
\newcommand{\pa}{\relax \ifmmode {\mbox Pa}\alpha\else Pa$\alpha$\fi}
\newcommand{\hb}{\relax \ifmmode {\mbox H}\beta\else H$\beta$\fi}
\newcommand{\ergs}{\relax \ifmmode {\,\mbox{erg\,s}}^{-1}\else \,\mbox{erg\,s}$^{-1}$\fi}
\newcommand{\me}{\relax \ifmmode {\,}^{-1}\else \,$^{-1}$\fi}
\newcommand{\degree}{\hbox{$^\circ$}}
\newcommand{\ergseccm}{erg\,sec$^{-1}$\,cm$^{-2}$}

\newcommand{\msun}{\relax \ifmmode {\,\mbox{M}}_{\odot}\else \,\mbox{M}$_{\odot}$\fi}

\newcommand{\cmtres}{\relax \ifmmode {\,\mbox{cm}}^{-3}\else \,\mbox{cm}$^{-3}$\fi}
\newcommand{\cmdos}{\relax \ifmmode {\,\mbox{cm}}^{-2}\else \,\mbox{cm}$^{-2}$\fi}
\newcommand{\cmseis}{\relax \ifmmode {\,\mbox{cm}}^{-6}\else \,\mbox{cm}$^{-6}$\fi}
\newcommand{\hi}{\relax \ifmmode {\mbox H\,{\scshape i}}\else H\,{\scshape i}\fi}
\newcommand{\arcminut}{$^{\prime}$}

\begin{document}     

%\slugcomment{submitted to ApJ}
%-----------------------------------------------------------------------------%
\title{Star Formation in Luminous HII regions in M33}
%-----------------------------------------------------------------------------%
\author{M\'onica Rela\~no\altaffilmark{1} and Robert C. Kennicutt, Jr.\altaffilmark{1}
}

\altaffiltext{1}{Institute of Astronomy, University of Cambridge,
  Madingley Road, Cambridge CB3 0HA, UK}

%-----------------------------------------------------------------------------%
\begin{abstract}
%-----------------------------------------------------------------------------%
 We present a multiwavelength (ultraviolet, infrared, optical and CO) study of a set of luminous \hii\ regions in M33: NGC~604, NGC~595, NGC~592, NGC~588 and IC131. We study the emission distribution in the interiors of the \hii\ regions to investigate the relation between the dust emission at 8\,\mi\ and 24\,\mi\ and the location of the massive stars and gas. We find that the 24\,\mi\ emission is closely related to the location of the ionized gas, while the 8\,\mi\ emission is more related to the boundaries of the molecular clouds consistently with its expected association with photodissociation regions (PDRs). Ultraviolet emission is generally surrounded by the \ha\ emission. For NGC~604 and NGC~595, where CO data are available, we see a radial gradient of the emission distribution at the wavelengths studied here: from the center to the boundary of the \hii\ regions we observe ultraviolet, \ha, 24\,\mi, 8\,\mi\ and CO emission distributions. We quantify the star formation for our \hii\ regions using the integrated fluxes at the set of available wavelengths, assuming an instantaneous burst of star formation. We show that a linear combination of 24\,\mi\ and \ha\ emission better describes the star formation for these objects than the dust luminosities by themselves. For NGC~604, we obtain and compare extinction maps derived from the Balmer decrement and from the 24\,\mi\ and \ha\ emission line ratio. Although the maps show locally different values in extinction, we find similar integrated extinctions derived from the two methods. We also investigate here the possible existence of embedded star formation within NGC~604.  
\end{abstract}                                          


%-----------------------------------------------------------------------------%
\section{Introduction}
%-----------------------------------------------------------------------------%
After the launch of the Spitzer Space Telescope and the Galaxy Evolution Explorer (GALEX) satellites the study of the star formation rate (SFR) in galaxies has improved considerably. The 24\,\mi\ and 8\,\mi\ Spitzer bands have been proposed as new SFR indicators in galaxies of different types (e.g. Calzetti et al. 2005; 2007, Alonso-Herrero et al. 2006). While the classical \ha\ and far-IR (FIR) emissions are directly linked to the star formation process, the \ha\ luminosity is produced by the recombination of photoelectrons and the FIR luminosity is produced by the amount of stellar light absorbed by dust, the consideration of both 24\,\mi\ and 8\mi\ emissions as SFR tracers relies on the observational correlation between the integrated luminosity at these two wavelengths and the 
extinction--corrected \ha\ luminosity. The relation has been found to hold in a statistically significant number of \hii\ knots in different galaxies. Calzetti et al. (2005) found a correlation between the 24\,\mi\ and the extinction--corrected 
\pa\ luminosities for the central \hii\ emitting knots in M51, which holds over more than two orders of magnitude in luminosity. A similar correlation was found for the \hii\ regions in M81, but with higher dispersion in the low-luminosity range and a larger range of dust opacities in those objects (P\'erez-Gonz\'alez et al. 2006). These correlations were also confirmed for Luminous Infrared Galaxies (LIRGs) and Ultraluminous Infrared Galaxies (ULRIGs) (Alonso-Herrero et al. 2006). The 8\,\mi\ emission was also studied as a star formation tracer in M51 and M81, but higher dispersion was found in the correlation of the 8\,\mi-\ha\ luminosities of \hii\ regions in these galaxies. A more complete study by Calzetti et al. (2007) involving a large galaxy sample shows the ability of the 24\,\mi\ emission to trace the star formation in galaxies of different types and metallicities (see also Wu et al. 2005 and Rela\~ no et al. 2007). These relations between dust emission and ionizing luminosity strictly hold only in very dusty \hii\ regions, where most of the stellar luminosity is reprocessed by dust. For \hii\ regions with a wider range of dust opacity, Kennicutt et al. (2007) show that a linear combination of the 24\,\mi\ emission, tracing the obscured star formation, and the observed \ha\ luminosity, which would trace the un-absorbed star formation, correlates better than other SFR tracers with the extinction--corrected \ha\ luminosity. This correlation has been confirmed in a large sample of \hii\ regions by Calzetti et al. (2007) and more recently for galaxies by Zhu et al. (2008).

The previously mentioned studies assume a direct correspondence between the 24\,\mi\ and \ha\ emissions. But although a considerable analysis has been carried out from a statistical point of view, little has been done to corroborate the spatial correlation of the emissions proposed as tracers of the star formation, and the geometries of the gas and dust relative to the 
position where  the stars actually form within \hii\ emitting knots. Using spatially resolved observations we are able to test the assumptions upon which the previous studies rest. The study of the emission distribution within an \hii\ region at the wavelengths that trace the SFR and the relation with the classical components of the \hii\ regions (central OB stars, ionized gas, photodissociation region (PDR) and molecular gas) offer an opportunity to test the hypotheses that are assumed in the statistical studies. Such an analysis, carried out in a set of giant \hii\ regions at different evolutionary stages, can help to test the basis for the results derived from the previous statistical correlations.  

M33 is an especially appropriate object because it offers a sample of giant \hii\ regions spanning wide ranges in luminosity and size. The distance of the galaxy (840~kpc; Freedman et al. 1991) allows an intermediate spatial resolution at the infrared wavelength range (at the distance of M33, the Multiband Imaging Photometer (MIPS) has a linear resolution of $\sim$20\,pc at 24\,\mi, which is $\sim$10 times smaller than the typical sizes of giant \hii\ regions). Among the brightest \hii\ regions in M33 there is a wide range of properties which allows us to select a complete set of giant \hii\ regions covering different luminosities and evolutionary stages. For example, IC131 contains a compact \hii\ region, IC131-West, and NGC~604 is a more evolved and dispersed high luminous \hii\ region.   

NGC~604 is of special interest. It is the brightest luminous \hii\ region in M33 and the second most luminous nearby \hii\ region after 30 Doradus in the Large Magellanic Cloud. Its ionized gas shows a complex morphology of cavities, shells and filaments that testify to the large amount of kinetic energy involved in the interior of the region. Relevant to this paper, the stellar content has been analyzed with filter imaging (e.g., Hunter et al. 1996, Drissen at al. 1993) and spectroscopically (e.g., Pellerin 2006, Terlevich et al. 1996, Gonz\'alez-Delgado et al. 2000); the ionized gas emission and dust extinction have been recently studied with Hubble Space Telescope (HST) images by Ma\' iz-Apell\' aniz et al. (2004). At longer wavelengths CO molecular gas has been observed by Wilson \& Scoville (1992) and radio emission has been analyzed by Churchwell \& Goss (1999). The size of the region and its evolutionary stage  make it very suitable for analyzing the correspondence of the IR Spitzer emission and the location of knots of star formation within the \hii\ region.  

We study a set of luminous \hii\ regions in the nearby galaxy M33 combining data at different wavelengths: 24\,\mi, 8\,\mi\ from Spitzer, $\sim$ 154\,nm, $\sim$ 232\,nm from GALEX, \ha\ and R-Band ground based data and \ha\ and \hb\ HST data. For two of the most luminous \hii\ regions in the sample we analyze CO molecular data.  We study the emission distribution at different wavelengths within the \hii\ regions in order to better understand the empirical correlations of the new proposals of star formation tracers in local environments. The rest of the paper is organized as follows. In \S 2 we explain the set of observations we analyze here, in \S 3 we describe the emission distribution within our set of \hii\ regions, \S 4 is devoted to obtain integrated fluxes for each \hii\ region and to convert them into SF measurements. We study NGC~604 in depth in \S 5, we summarize our conclusions and discuss the results in \S 6.  

%-----------------------------------------------------------------------------%
\section{Sample and Data}
\label{S2}
\begin{figure*}
\epsscale{0.9}
\includegraphics[width=0.5\textwidth]{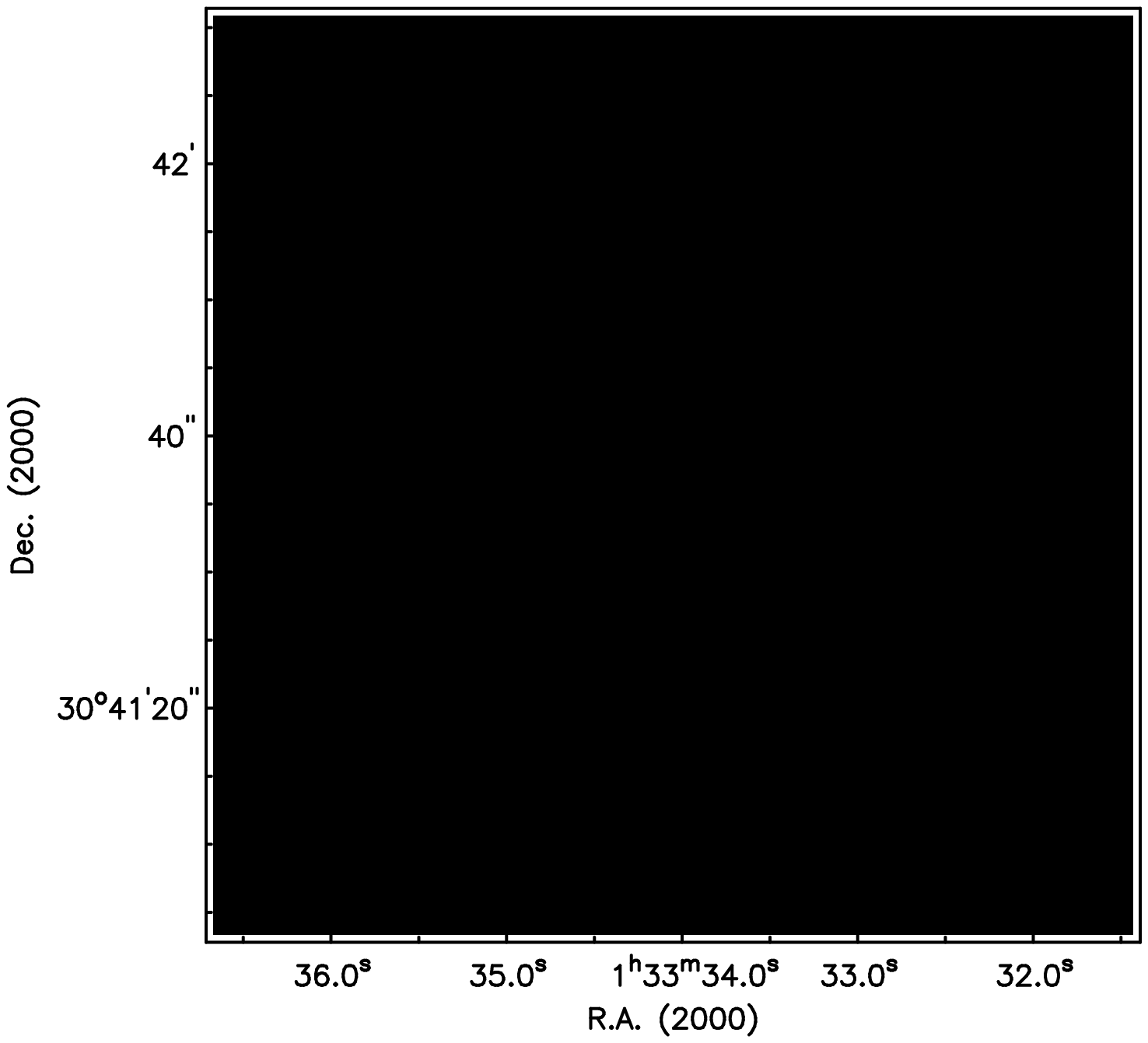}
\includegraphics[width=0.5\textwidth]{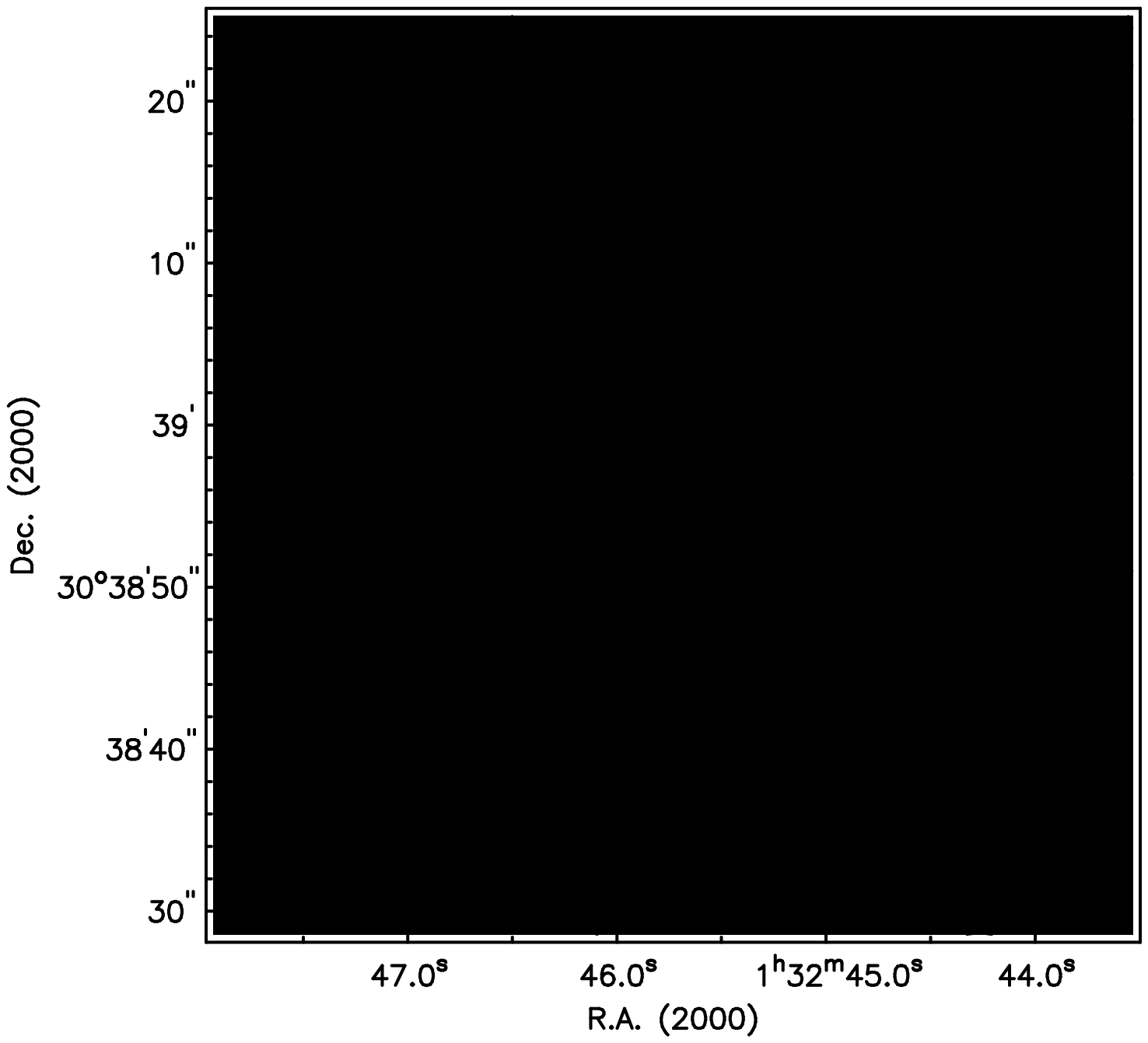}  
\includegraphics[width=0.5\textwidth]{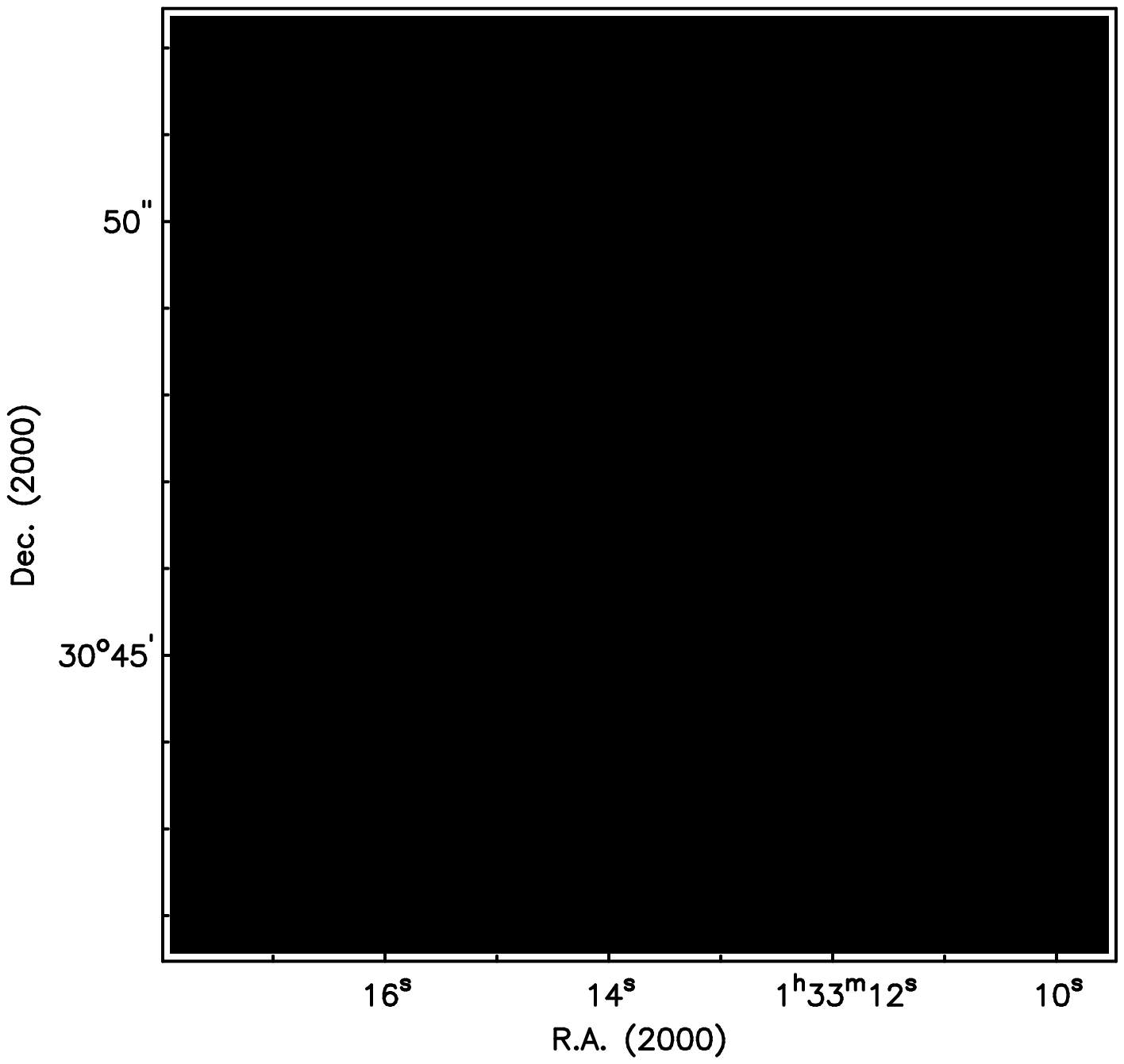} \hfill
\includegraphics[width=0.5\textwidth]{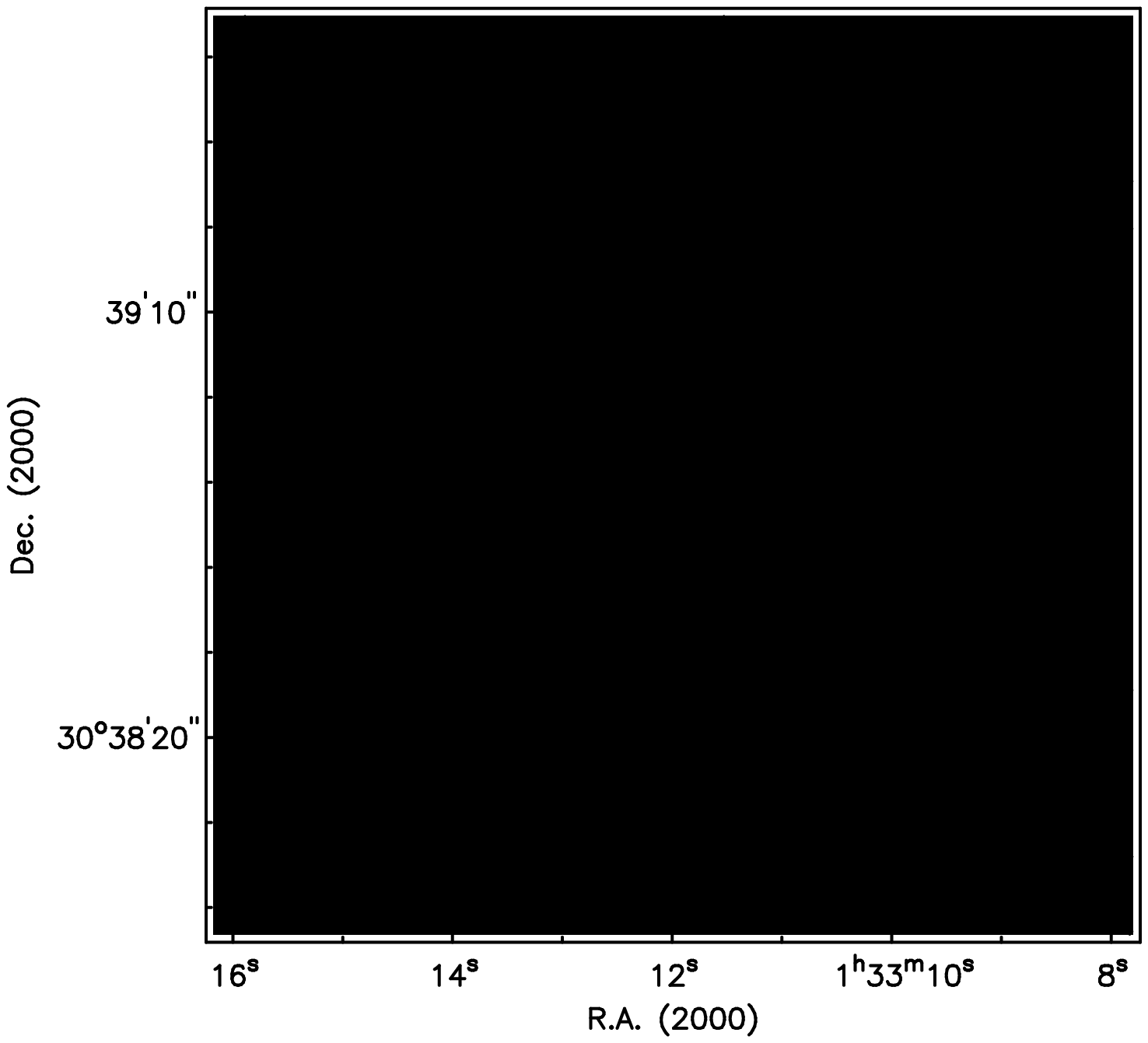}
\caption{Continuum-subtracted \ha\ images of our set of \hii\ regions in M33: NGC~595 (upper left), NGC~588 (upper right), IC131 and IC131-West (lower left) and NGC~592 (lower right) with 24\,\mi\ emission contours overlaid. The intensity contours are at (2, 5, 10, 20, 40, 60, 80, 95)\% of the 24\,\mi\ maximum intensity in each region. A 2\% contour level corresponds to a range of (2-6)$\sigma$, depending on the region (3$\sigma$=2.0$\times$10$^{-14}$~\ergseccm\ for the 24\,\mi\ image). A 3$\sigma$ value in the continuum-subtracted \ha\ image corresponds to 2.0$\times$10$^{-17}$~\ergseccm.}
\label{fig1}
\end{figure*}

%-----------------------------------------------------------------------------%
We have selected a sample of bright \hii\ regions in M33: NGC~588, NGC~592, NGC~595 and NGC~604, IC131 and IC131-West (the western radio component of IC131, see Viallefond et al. 1983). All of them show high \ha\ luminosity over the range 
2$\times$10$^{38}$--3$\times$10$^{39}$\ergs. We have chosen large \hii\ regions whose sizes allow us to spatially resolve their structure at the wavelengths we study here (the spatial scale of the observations presented here range from $\sim$1\,pc for the HST data to $\sim$32\,pc for the CO observations). The \hii\ regions of our sample are at  different evolutionary stages -from the compact IC131-West to the more evolved NGC~604- and show different \ha\ morphologies -from the open \ha\ shell structure of NGC~595 to a well defined complete \ha\ shell shown by NGC~588 or the multiple arcs and filaments of NGC~604 (see Figures.~\ref{fig1} and~\ref{fig2}). There are available data from UV to the IR for our set of \hii\ regions and for two of them, NGC~604 and NGC~595 we are able to compare with CO intensity maps. Finally, for NGC~604 we also analyze \ha\ and \hb\ images from the HST Data Archive. In this paper we have used a data combination from Spitzer (Werner et al. 2004), GALEX (Martin et al. 2005) and ground-based \ha\ observations from the Local Group Galaxies Survey (LGGS) (Massey et al. 2006, 2007). For NGC~604 and NGC~595 we have compared these data with CO intensity maps from Wilson \& Scoville (1992).

\subsection{IR Data: Spitzer}
The Infrared Array Camera (IRAC, Fazio et al. 2004) on Spitzer is an imaging camera operating at four  ``channels'' (3.6,4.5,5.8 and 8\,\mi). The field of view is 5.$^{\prime}$2$\times$5.$^{\prime}$2 in the full array readout mode (256$\times$256 pixels). The plate scale is 1.$^{\prime\prime}$2/pixel. M33 was observed six times using IRAC under Program ID 5 (PI: R. Gehrz). The Basic Calibrated Data (BCD)  created by the Spitzer Data Center (SSC) pipeline version S14.0.0 were taken from the Spitzer Data Archive and assembled with MOPEX version 16.3.7\footnote{http://ssc.spitzer.caltech.edu/postbcd/mopex.html}. The images were oriented and aligned using common point sources and background subtraction in all the images was done before combining them together. Since the 8\,\mi\ image is dominated by the PAH (Polycyclic Aromatic Hydrocarbon) emission and most of the 3.6\,\mi\ emission can be assumed to have a photospheric stellar origin (Helou et al. 2004), we have used the 3.6\,\mi\ image to estimate the stellar contribution of the 8\,\mi\ flux. We have followed the method described in Helou et al. (2004) (see also Calzetti et al. 2007) and generated scaled 3.6\,\mi\ images with scale factors within a range of 0.22-0.28. We found that the scale factor which gave least residuals after subtracting the stellar component from the 8\,\mi\ image was 0.24. We estimated an uncertainty of $\la$1\%  in the integrated 8\,\mi\ fluxes of our \hii\ region sample for a change in the scale factor of 0.04.

Observations of M33 at 24\,\mi\ were obtained with the Multiband Imaging Photometer (MIPS) in the Scan-Mode on Spitzer (Rieke et al. 2004). The field of view is 5\arcminut$\times$5\arcminut\ and the resolution is 5.$^{\prime\prime}$7. We retrieved the BCDs of Program ID 5 corresponding to position 2 (RA(2000)=1h34m10.00s, DEC(2000)=30d47m02.0s), and position 3 (RA(2000)=1h33m30.00s, DEC(2000)=30d32m11.0s), which cover the portion of the galaxy disk where our set of \hii\ regions are located. The BCD data were created using the SSC pipeline software version S16.1.0 and the data were assembled using Mopex version 16.3.7, preserving the original pixel size of 2.$^{\prime\prime}$45. For each pointing the images were oriented and aligned using several point sources and a sky level was obtained and subtracted for each image.  After the subtraction of a constant sky level, a small gradient in the background was still seen in all the images. Prior to combining the images the background gradient was removed using the task imsurfit in IRAF\footnote{IRAF is distributed by the National Optical Astronomy Observatory, which is operated by the Association of Universities for Research in Astronomy (AURA), Inc., under cooperative agreement with the National Science Foundation.}. Although MIPS (Multiband Imaging Photometer) provides observations at longer wavelengths (70\,\mi\ and 160\,\mi), the spatial resolution achieved at these wavelengths is low for the purposes of this study ($\sim$60\,pc and $\sim$150\,pc for 70\,\mi\ and 160\,\mi, respectively). For this reason, we have not included them in the analysis presented here.

\subsection{UV Data: GALEX}
                
The Nearby Galaxies Survey (NGS) includes observations of nearby galaxies of different types and environments, M33 among others (Bianchi et al. 2003a, 2003b). The far-ultraviolet (FUV) (1344-1786\AA, $\lambda _{\rm eff}$=1539\AA) and near-ultraviolet (NUV) (1771-2831\AA, 
$\lambda _{\rm eff}$=2316\AA) data of M33 were taken from the GR2 data release of the GALEX-Nearby Galaxies Survey, which is available to the public via the Multimission Archive at the Space Telescope Science Institute (MAST)\footnote{http://galex.stsci.edu/GR2/}. The calibrated images obtained from the Data Archive were used here. They have a resolution of 4.$^{\prime\prime}$2 and 5.$^{\prime\prime}$3 in FUV and NUV, respectively. Data reduction and image calibration for the
GR2 data release is explained in Morrissey et al. (2007).

\subsection{Ground-based \ha\ Imaging}

We use \ha\ emission observations provided by the KPNO/CTIO Local Group Galaxy Survey (LGGS) collaboration, 
which has already published UBVRI catalogs of stars in M31 and M33 (Massey et al. 2006), 
and also provides images at \ha, [\sii] and [\oiii] (Massey et al. 2007). 
The M33 data were taken with the Mosaic CCD camera at the prime focus of the 4~m Mayall Telescope, the observations have a plate 
scale of 0.$^{\prime\prime}$27 pixel\me\ and a spatial resolution of 0.$^{\prime\prime}$8. We have retrieved two images of M33 from the final reduced data available at the NOAO (National Optical Astronomy Observatory) Science Archive: one obtained with 
the narrow-band \ha- k1009 (6575/81\AA) filter and the other with the broad band filter 
R Harris k1004 (6514/1511\AA) used for continuum subtraction. The \ha\ and continuum images were 
aligned using the positions of several field stars and aperture photometry of these stars was used to derive a scale factor 
of 0.41$\pm$0.05. We generated continuum-subtracted \ha\ images with scale factors in the range of 0.36-0.46. A scale factor of 0.44 was finally chosen after careful inspection of the residuals in the continuum subtraction process. 
An uncertainty of 1-3\% in the integrated \ha\ fluxes was estimated for a change in the 
scale factor of 0.05. Contributions of [\nii]$\lambda$6548,6584 emission lines to the integrated fluxes were obtained using the 
transmission curve of the \ha-k1009 filter and the expected [NII]/\ha\ emission 
line ratios for each \hii\ region in our sample (Bosch et al. 2000 and V\'ilchez et al. 1988). We applied a calibration 
factor of 1.79$\times$10$^{-16}$\ergseccm\ given in Table~2 in Massey et al. (2007)\footnote{There is an error in the units of the \ha\ calibration factor 
listed in Table~2 of Massey et al. (2007) that 
we have checked with the authors of this paper (P. Massey priv. com.)} for the \ha\ emission line sources. This calibration was checked by comparing 
the integrated fluxes with previously reported \ha\ fluxes (see section~\ref{S4}).

\subsection{Hubble Space Telescope Wide-Field Planetary Camera 2 Imaging of NGC~604}

We retrieved \ha\ and \hb\ observations of NGC~604 from ID Programs 5773 and 9134 from the HST MAST Data Archive.  We selected images at three different filters: F656N, F487N to isolate the \ha\ and \hb\ emission and F547M for continuum subtraction. The observations for the 
F656N, F487N and F547M filters consist of 2$\times$1100~s, 3$\times$2700~s and 2$\times$500~s, respectively. The images were aligned and cosmic rays were rejected using STSDAS package {\it crrej} in IRAF. Wide-Field Planetary Camera 2 (WFPC2) mosaic images were then obtained using the task {\it wmosaic} in the STSDAS package. The resulting \ha, \hb\ and continuum (F547M) images were aligned using positions of stars in the images. Using aperture photometry for field stars, we obtained scale factors of 0.11$\pm$0.04 and 0.18$\pm$0.05 for \ha\ and \hb\ respectively. A detailed inspection of the images showed that the best scale factors were 0.14 and 0.16 for the  \ha\ and \hb\ images respectively. Changes in the scale factors of 0.04 and 0.05 give flux uncertainties of  
3\% and 7\% for \ha\ and \hb, respectively. The final resolution of the \ha\ and \hb\ images is $\sim$0.$^{\prime\prime}$2. We used the absolute photometric calibrations of the WFPC2 nebular filters to measure the integrated flux of the region. Corrections for the contamination of the [\nii]$\lambda$6548,6584 emission lines were performed for the \ha\ image and the final integrated flux was compared with previously reported fluxes from the literature (see \S 4).

\subsection{Data from Literature}

We also make use of data available in the literature at other wavelengths. We use interferometric 
CO observations of NGC~604 and NGC~595 with a synthesized beam of 
$\sim$7\arcsec$\times$8\arcsec\, as published by Wilson \& Scoville (1992). We also analyze the results of the stellar photometry 
of NGC~604 previously published by Hunter et al. (1996) and use the F555W image of this \hii\ region for a comparison with the CO data.

%-----------------------------------------------------------------------------%
\section{Emission Distribution Within the HII Regions}
\label{S3}
%-----------------------------------------------------------------------------%

The study of the components of the interstellar dust is improving considerably since the launch of Spitzer  (e.g. Draine \& Li 2007, Lebouteiller et al. 2007). The general 
model of the interstellar dust suggests the existence of three different components: large grains (emitting at wavelengths 
$\lambda>$ 50\,\mi), very small grains (VSG) (emitting at $\lambda>$ 10\,\mi) and PAH molecules with emission at concentrated 
features in the 3-15\,\mi\ region (D\'esert et al 1990; Draine 2003). In this general picture we would expect that the emission corresponding to the 24\,\mi\ MIPS band will be related to the emission of the VSG, while the 8\,\mi\ IRAC band will include part of the PAH emission features. The relation of the dust model components and the infrared (IR) emission in the IRAC and MIPS bands has been recently studied by Draine \& Li (2007). While the VSG are heated by single photons from the stellar radiation field producing a broad temperature distribution and are believed to be located close to the central stellar cluster in star forming regions (Cesarsky et al. 1996, Lebouteiller et al. 2007); the PAH grains can be of various sizes emitting at different bands (Draine \& Li 2001) and their emission is seen to peak in the interfaces between the \hii\ regions and the molecular clouds and PDRs (Cesarsky et al. 1996). 
 
In this section, we analyze the emission at 8\,\mi\ and 24\,\mi\ in the interior of a set of \hii\ regions (NGC~604, NGC~595, NGC~588, NGC~592 and IC131) 
and relate them to the UV and \ha\ emission observed in the \hii\ regions. NGC~604 and NGC~595, which are more extensively studied objects than the rest, 
will be discussed separately below (see \S 5).

\subsection{24\,\mi\ and \ha\ Emission}
\begin{figure*}
\includegraphics[width=0.5\textwidth]{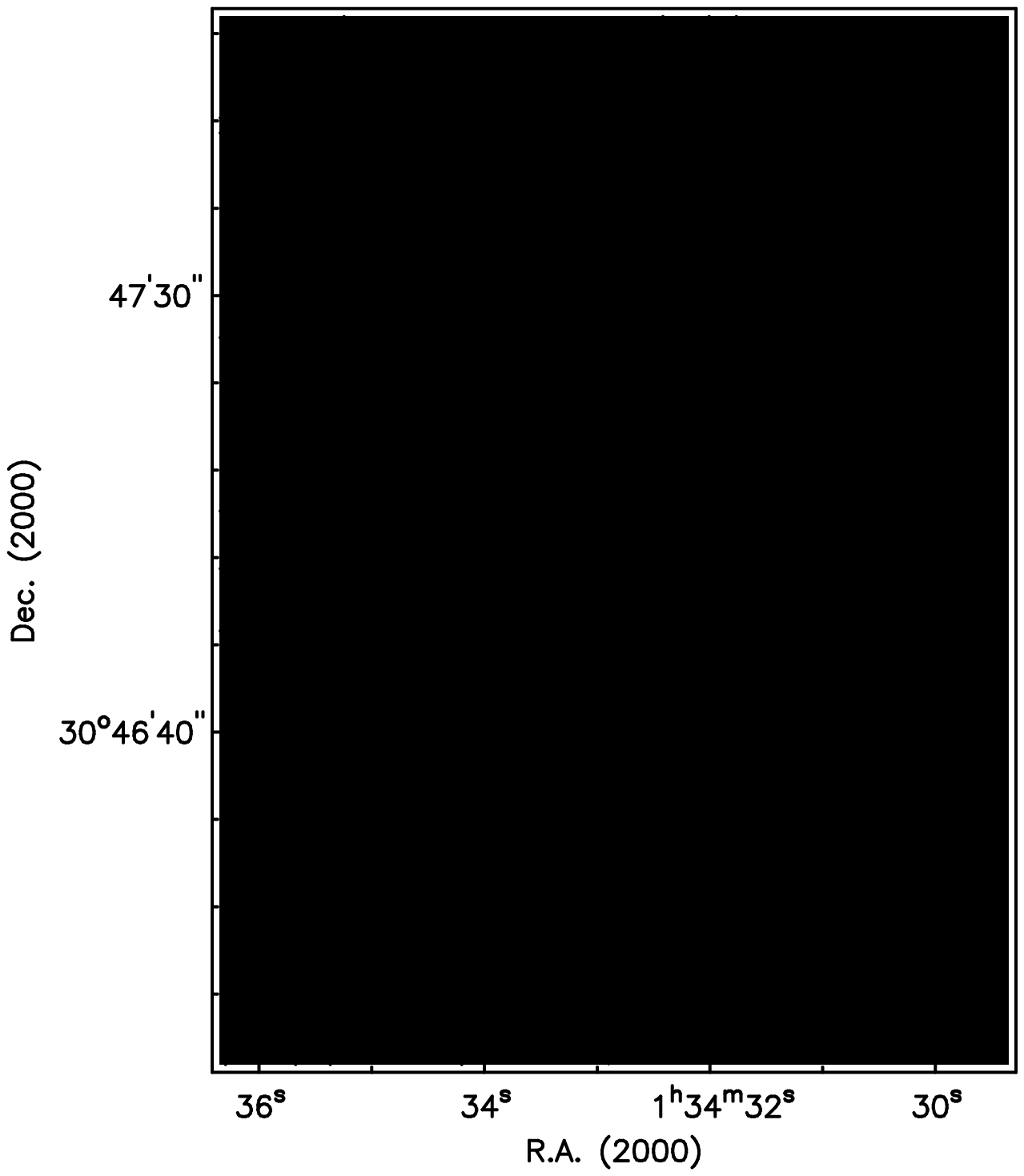}
\includegraphics[width=0.5\textwidth]{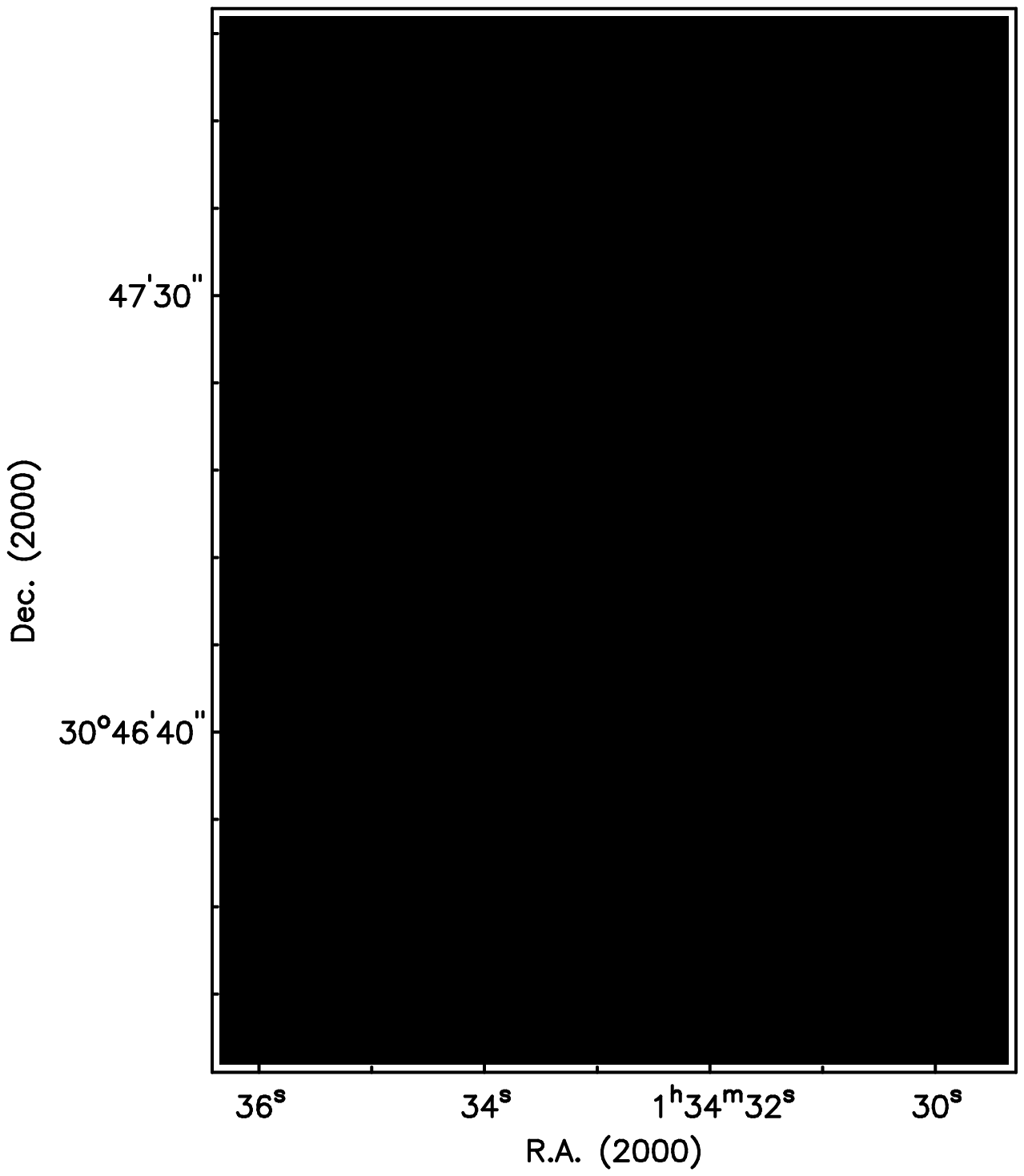}  
\includegraphics[width=0.5\textwidth]{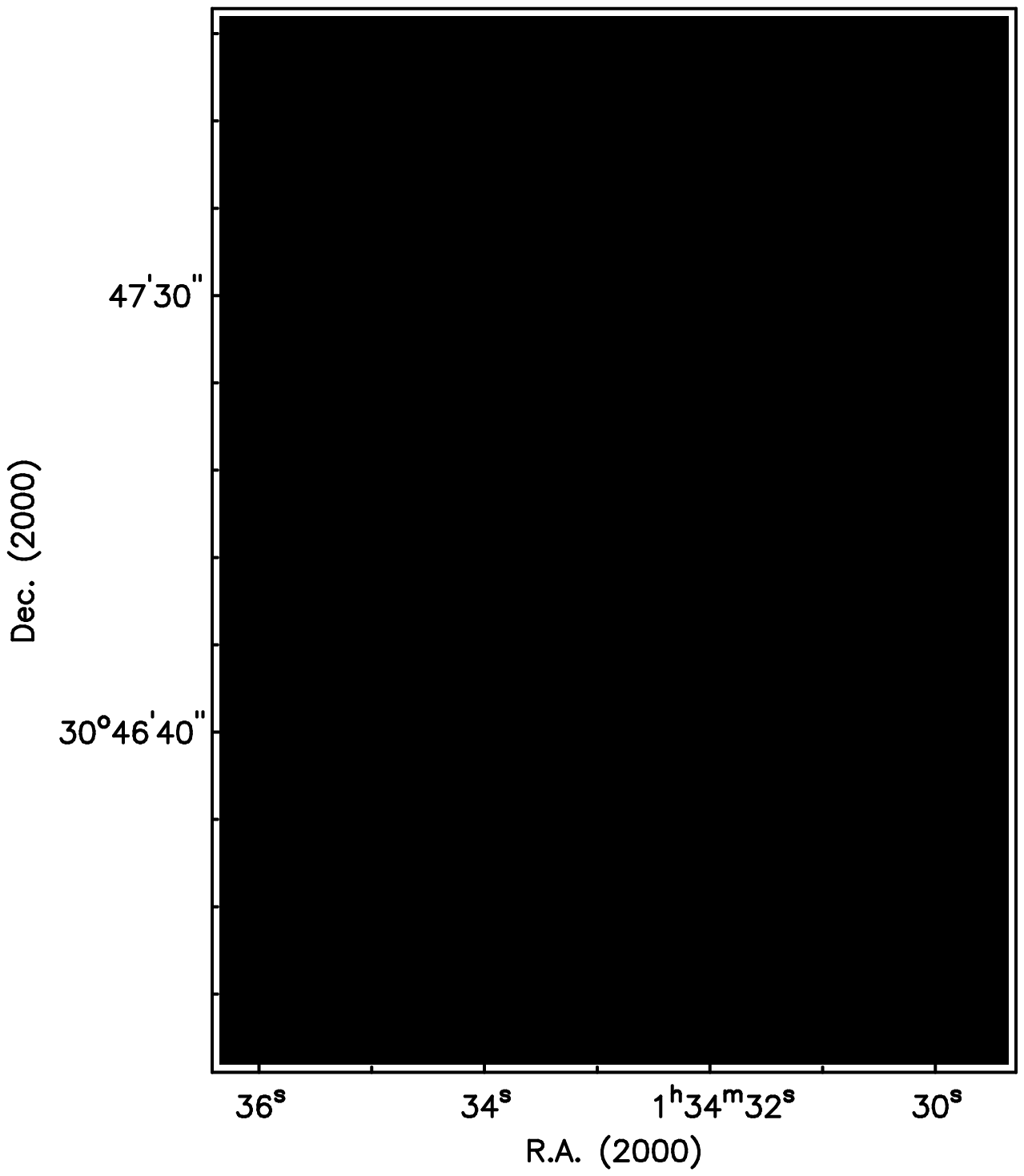} \hfill
\includegraphics[width=0.5\textwidth]{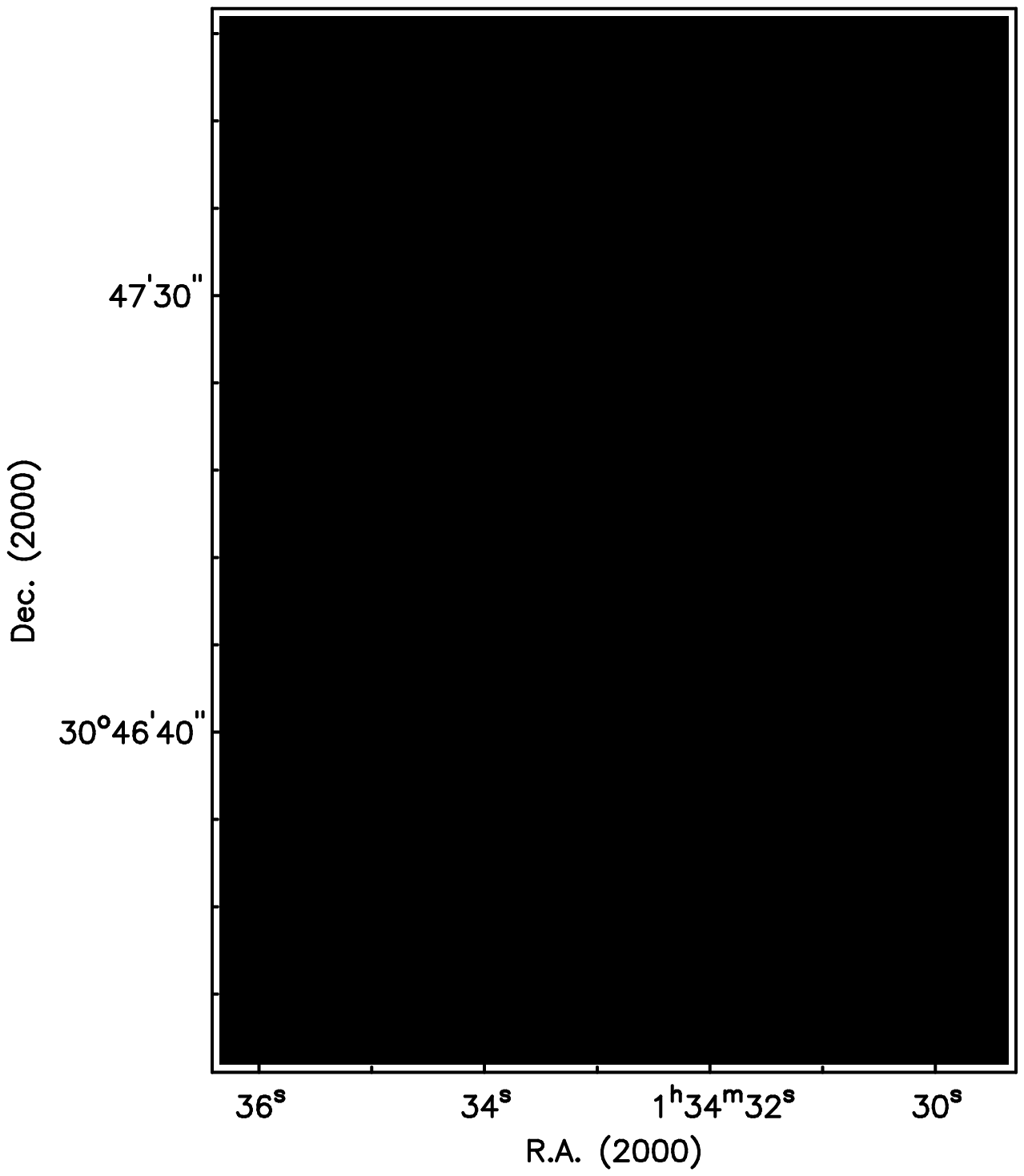}  
\caption{Upper: NGC~604 continuum-subtracted \ha\ image at high resolution with 8\,\mi\  (left) and UV (right) emission contours overlaid. The intensity contours are at (2, 5, 10, 20, 40, 60, 80, 95)\% of the maximum intensity at each wavelength (2\% contour level corresponds to 3$\sigma$ (9.6$\times$10$^{-15}$~\ergseccm\ ) and 20$\sigma$ (1.0$\times$10$^{-13}$~\ergseccm\ ) at 8\,\mi\ and UV, respectively).  A 3$\sigma$ value in the continuum-subtracted \ha\ image of NGC~604 corresponds to 4.2$\times$10$^{-18}$~\ergseccm. Lower left: Observed \ha\ luminosity of NGC~604 at 6\arcsec\ resolution with 24\,\mi\ emission contours overlaid. Lower right: Absorbed \ha\ luminosity for NGC~604 derived using the Balmer extinction map in Figure~\ref{fig8}-left with 24\,\mi\ emission contours overlaid. The intensity contours are at (2, 5, 10, 20, 40, 60, 80, 95)\% of the maximum 24\,\mi\ intensity, (2\% contour level corresponding to 10$\sigma$).}
\label{fig2}
\end{figure*}

We first analyze the spatial correspondence between the 24\,\mi\ and \ha\ emissions. In Figure~\ref{fig1} we show continuum-subtracted \ha\ images for the selected \hii\ regions 
with 24\,\mi\ contours overlaid. There is a clear spatial correlation between \ha\ and 24\,\mi\ emission in all \hii\ regions 
of our sample. The most luminous \ha\ knots correspond spatially to the most 
luminous ones at  24\,\mi\ and the \ha\ morphology is perfectly traced by the 24\,\mi\ emission 
at lower resolution: the most luminous knots at 24\,\mi\ and \ha\ are located 
at the same position within NGC~588, the same is seen in NGC~592, and 
IC131-West, a very concentrated \ha\ knot, shows the same highly concentrated morphology at 24\,\mi. The same spatial correspondence is seen for NGC~604 in Figure~\ref{fig2} (lower left panel), where the \ha\ emission at 6\arcsec\ resolution is compared to the 24\,\mi\ emission. In NGC~595 the agreement is not so good: the 24\,\mi\ emission peaks in between the two central \ha\ maxima, which could be due to a higher extinction at this particular position within the region (see Figure~18 in Bosch et al. 2002), but in general the 24\,\mi\ emission follows the same shell-like distribution as \ha\ in this region. The strong correspondence between the emission at both wavelengths confirms the general trend observed in other galaxies (e.g. Calzetti et al. 2005; P\'erez-Gonz\'alez et al. 2005, Alonso-Herrero et al. 2006, Prescott et al. 2007) and in nearby \hii\ regions (Churchwell et al. 2006; Watson et al. 2008). 

The good spatial correlation between 24\,\mi\ and \ha\ emissions shows that the dust emitting at 24\,\mi\ would be 
predominantly heated by the emission coming from the OB stars within the \hii\ regions.
In this situation, the 24\,\mi\ emission radiated by the dust, that has previously absorbed the light 
coming from the stars, would be more related to the absorbed \ha\ luminosity than to the observed \ha\ luminosity in the region. In order to test this hypothesis we have produced a map of absorbed \ha\ luminosity of 
NGC~604 (Figure~\ref{fig2}, right panel) using the following method. We have obtained a map of the gas extinction using HST continuum subtracted \ha\ and \hb\ images (the derivation of this map will be explained below in \S~\ref{S5}), then the extinction map was 
used to derive an extinction--corrected \ha\ luminosity map of the region. Assuming a color excess of E(B$-$V) = 0.07~mag (van den Bergh 2000) and Cardelli et al. (1989) extinction law with $\rm R_{\rm V}$=3.1, we derive a contribution of A(\ha)=0.17 for the foreground Galactic extinction. The \ha\ luminosity absorbed in the \hii\ region will be the difference between the total extinction--corrected \ha\ luminosity and the \ha\ luminosity corrected for the foreground Galactic extinction. The result is shown in the right panel in Figure~\ref{fig2} with 24\,\mi\ emission contours overlaid. In spite of the relatively low extinction in NGC~604 we can clearly see that the absorbed \ha\ luminosity (right panel) correlates better  with the 24\,\mi\ emission than the observed \ha\ luminosity (left panel). This result qualitatively confirms previous results showing that the 24\,\mi\ is directly linked to the {\it extincted} star formation (Kennicutt et al. 2007; Calzetti et al, 2007; P\'erez-Gonz\'alez et al. 2006). We further quantitatively investigate this relationship in \S~\ref{S4.1}.

\subsection{8\,\mi\ Emission}
\begin{figure*}
\includegraphics[width=0.5\textwidth]{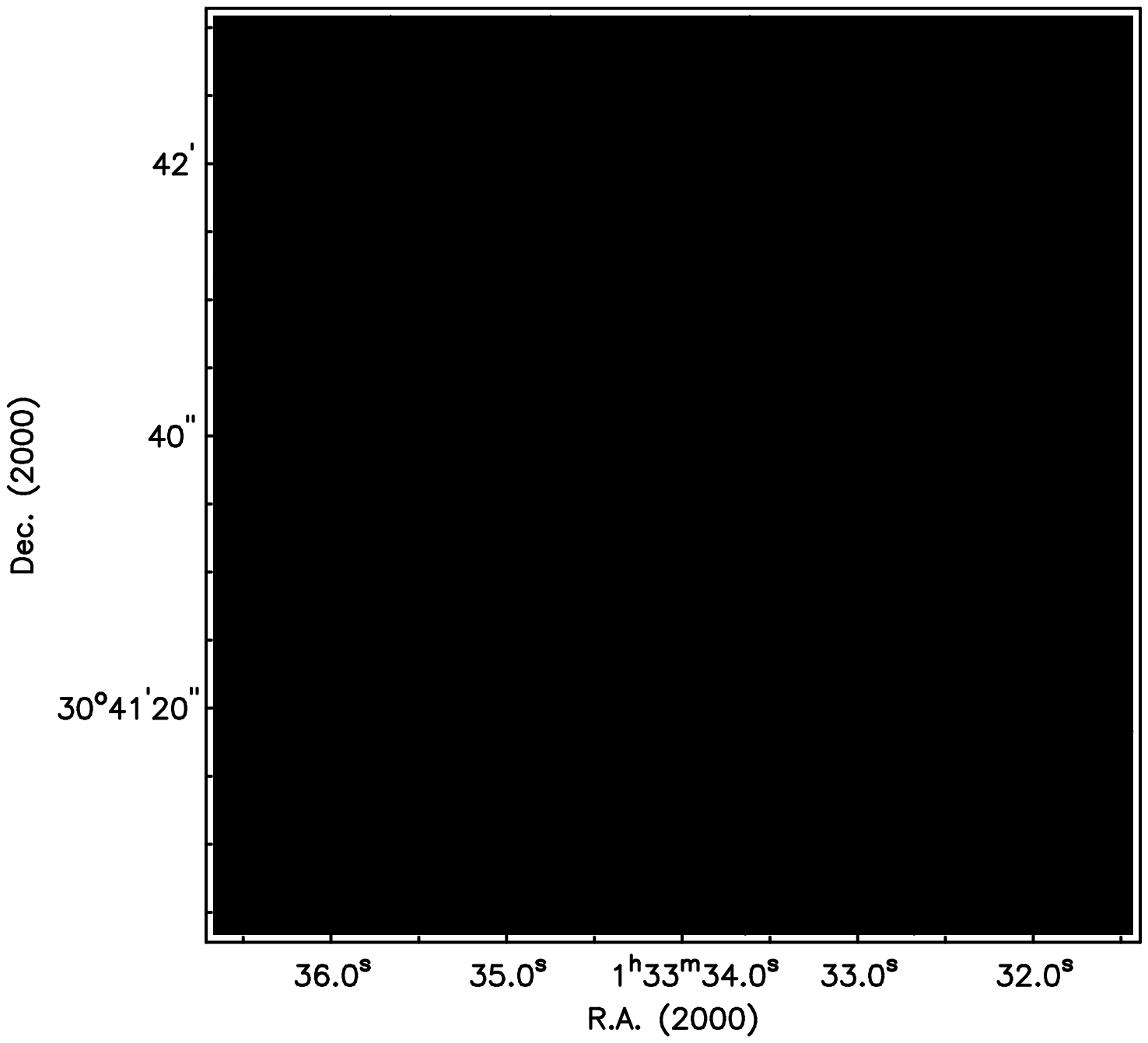}
\includegraphics[width=0.5\textwidth]{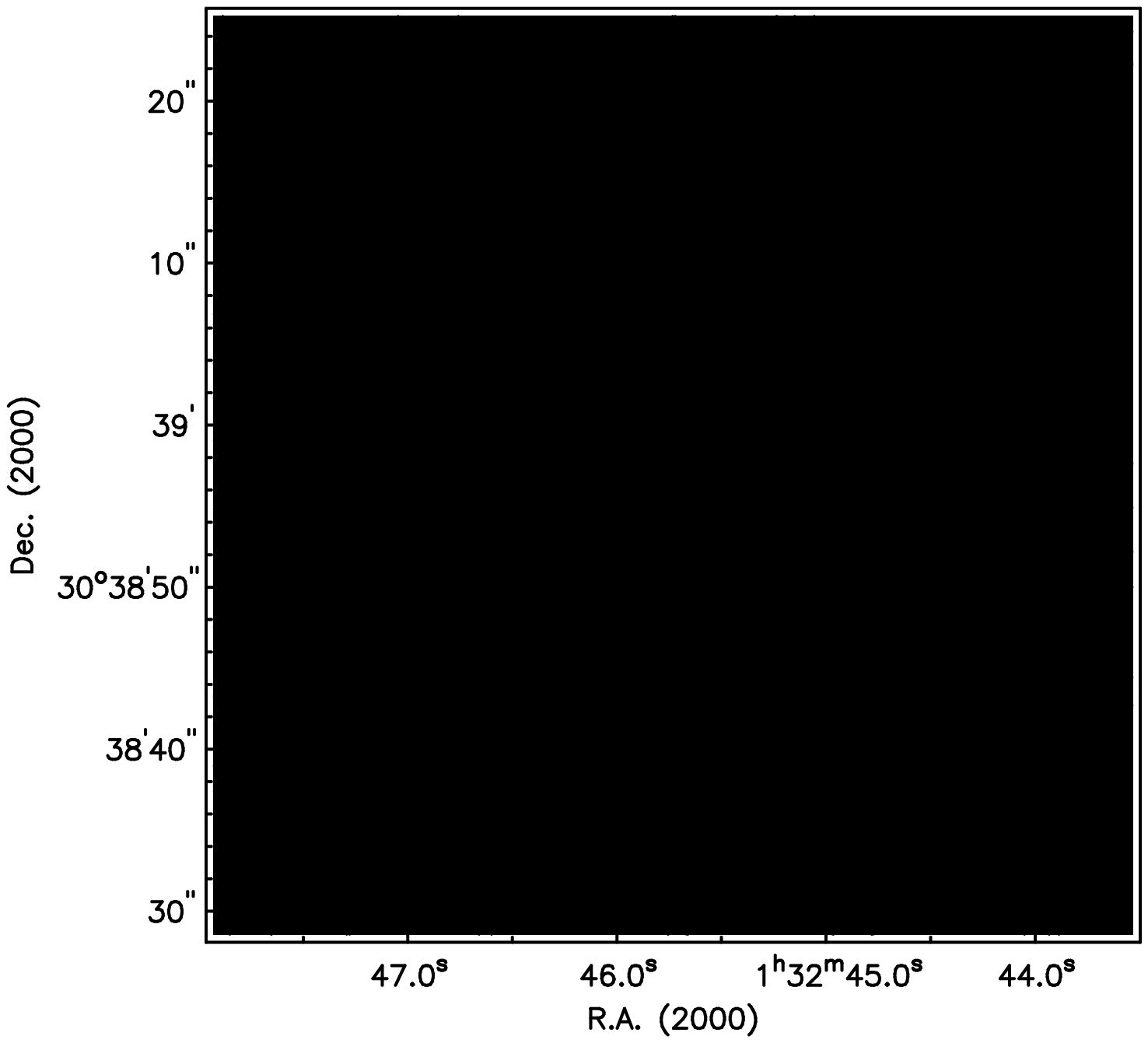}
\includegraphics[width=0.5\textwidth]{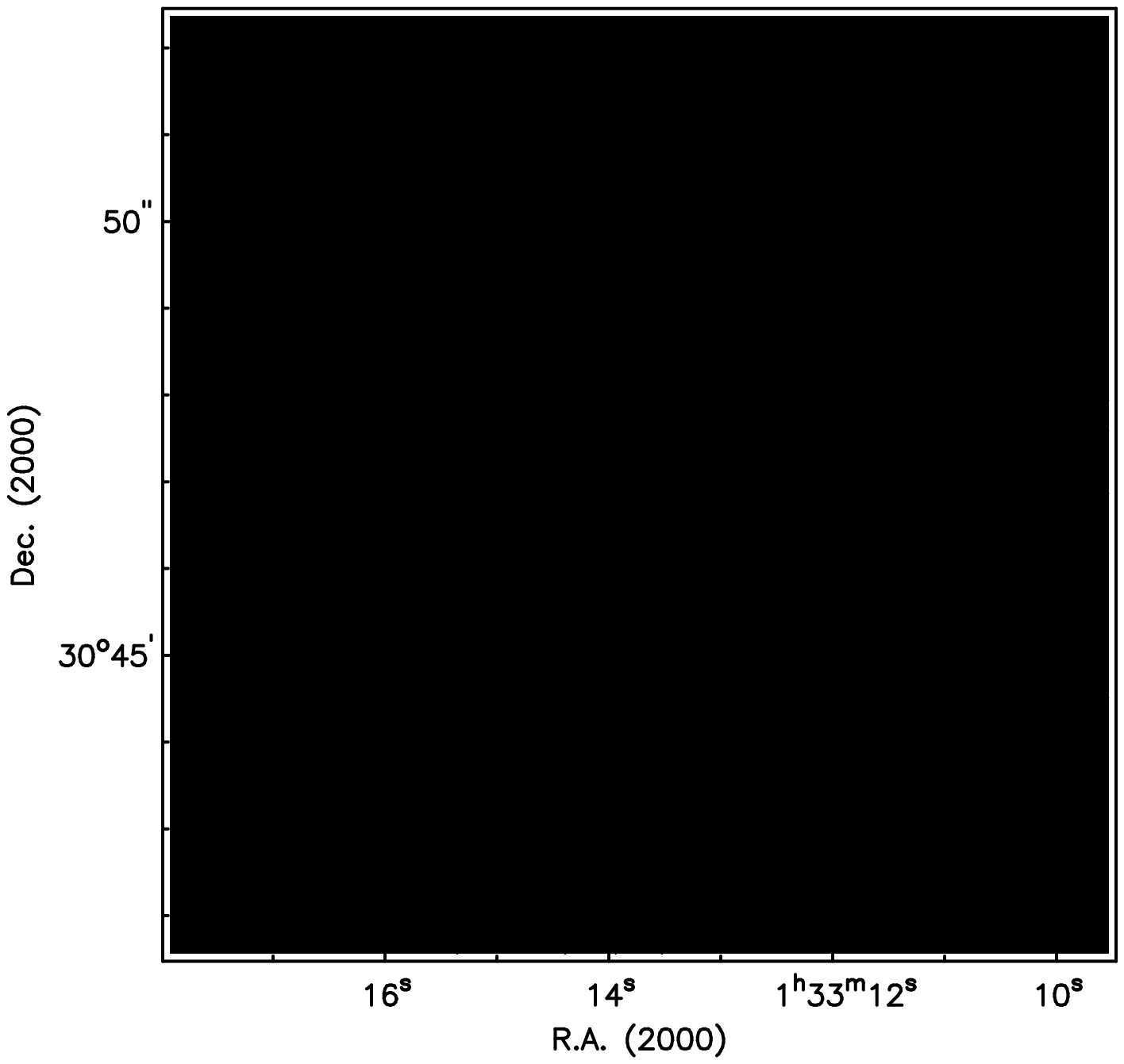} \hfill
\includegraphics[width=0.5\textwidth]{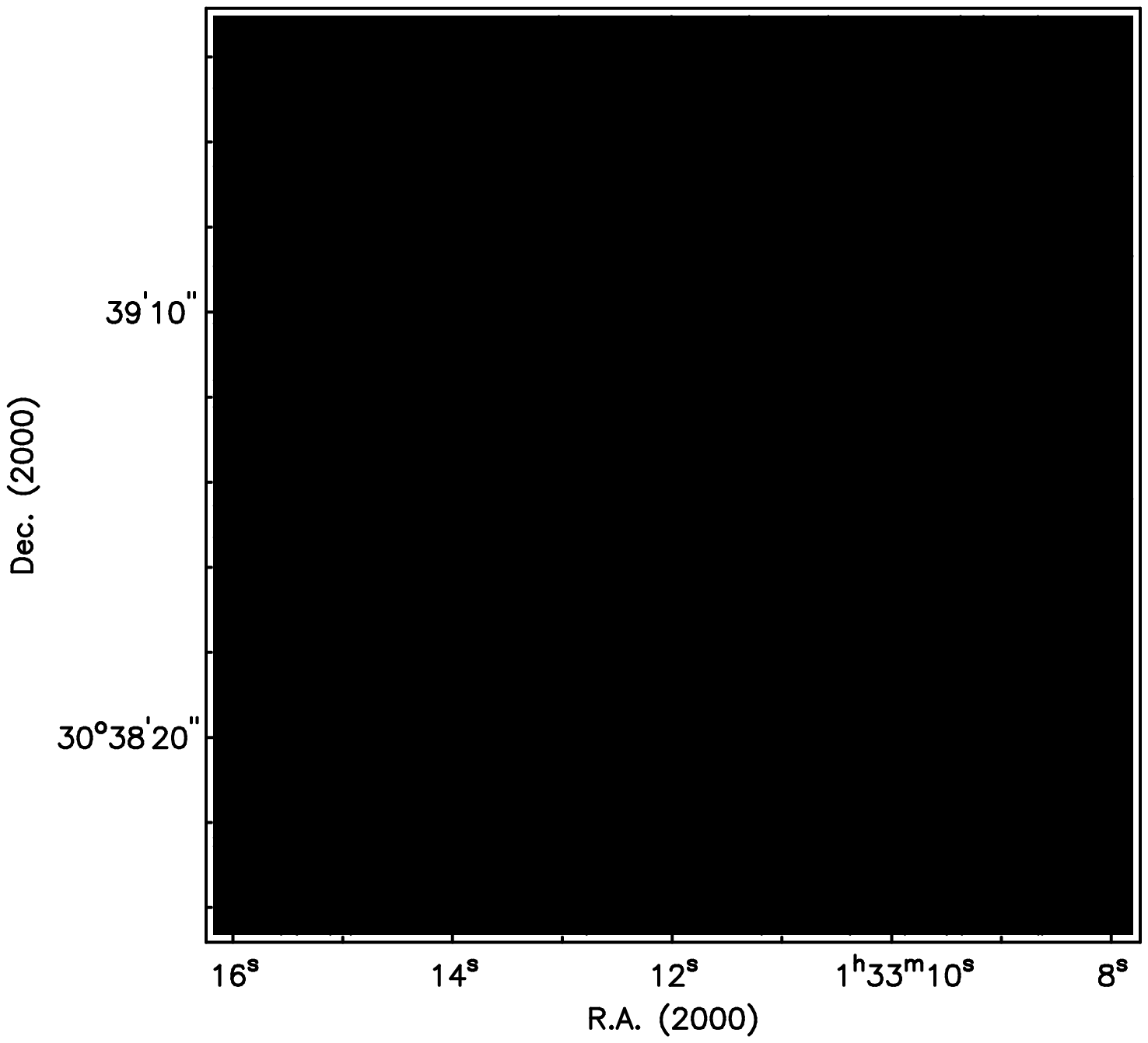}
\caption{Continuum-subtracted \ha\ images of our set of \hii\ regions in M33: NGC~595 (upper left), NGC~588 (upper right), IC131 and IC131-West (lower left) and NGC~592 (lower right) with 8\,\mi\ emission contours overlaid. 
The intensity contours are at (2, 5, 10, 20, 40, 60, 80, 95)\% of the maximum intensity.  A 2\% contour level corresponds to a range of (2-6)$\sigma$, depending on the region.}
\label{fig3}
\end{figure*}

The 8\,\mi\ emission shows differences from the 24\,\mi\ emission distribution in the interiors of our set of \hii\ regions. In Figure~\ref{fig3} we show the continuum-subtracted \ha\ images with 8\,\mi\ emission contours overlaid. In the outskirts of the \hii\ regions the 8\,\mi\ emission traces the filamentary structure, delineating in some cases the \ha\ filaments and shells (e.g. in NGC~595 and NGC~588 and NGC~604 in Figure~2, upper left panel). In the central part of the \hii\ regions the 8\,\mi\ maxima is in general displaced from the \ha\ maxima (e.g. NGC~595 and western part of NGC~604), while the maxima at 24\,\mi\ coincide with \ha\ maxima. The morphology suggests that 
the 8\,\mi\ emission is more related to the PDR than to the location of the ionizing stars or the ionized gas. Differences between 24\,\mi\ and 8\,\mi\ emissions within extragalactic \hii\ regions were also suggested by Helou et al. (2004) for NGC~300, but the linear resolution of their data (a factor of 2.5 lower than ours) did not allow them to make a firm statement about this. Similar results related to the distribution of the 24\,\mi, 8\,\mi\ and \ha\ emission as these shown here for our set of \hii\ regions have been reported for Galactic \hii\ regions (e. g. Churchwell et al. 2006; Watson et al. 2008).

\subsection{UV Emission}
\begin{figure*}
\includegraphics[width=0.5\textwidth]{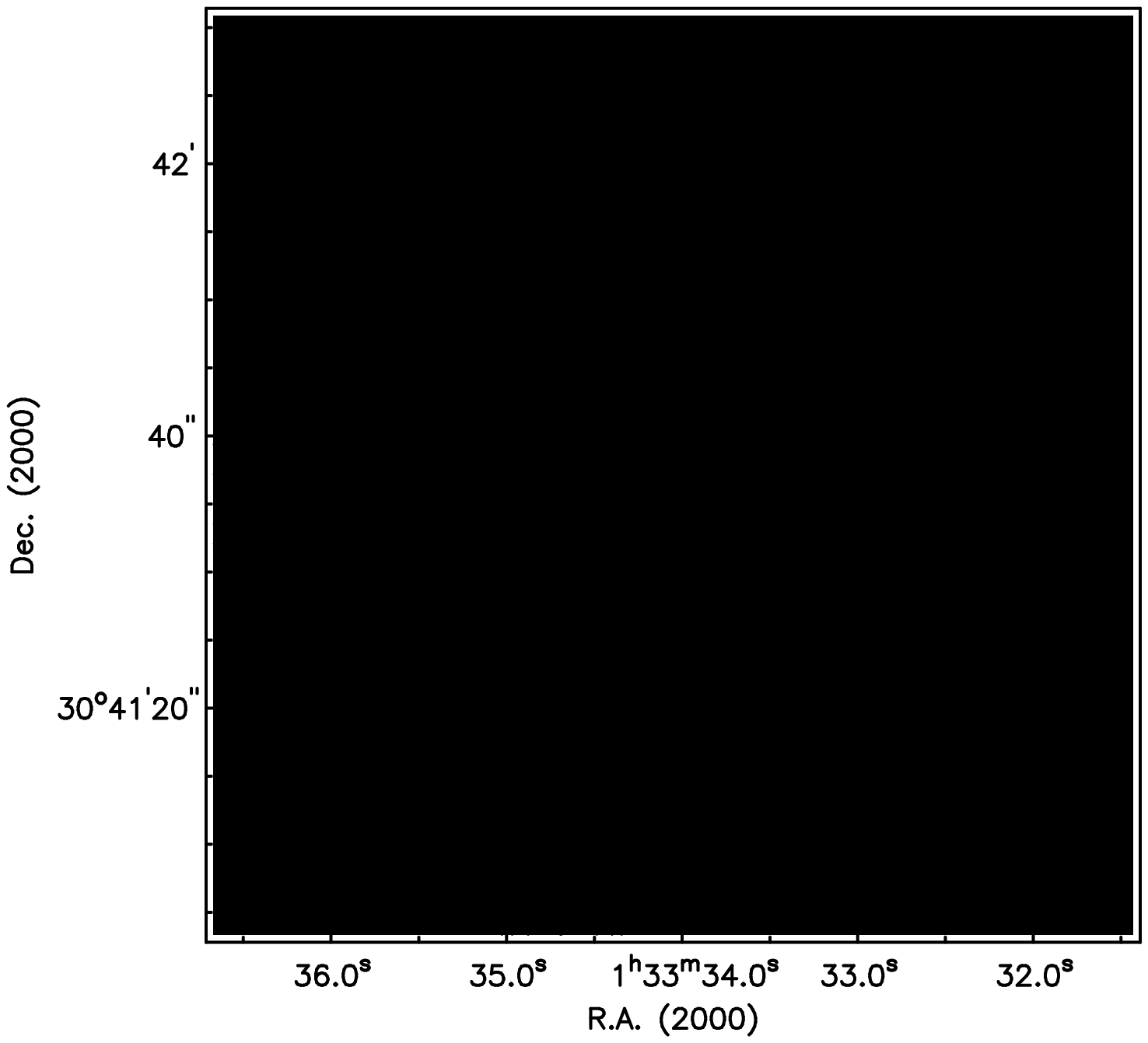}
\includegraphics[width=0.5\textwidth]{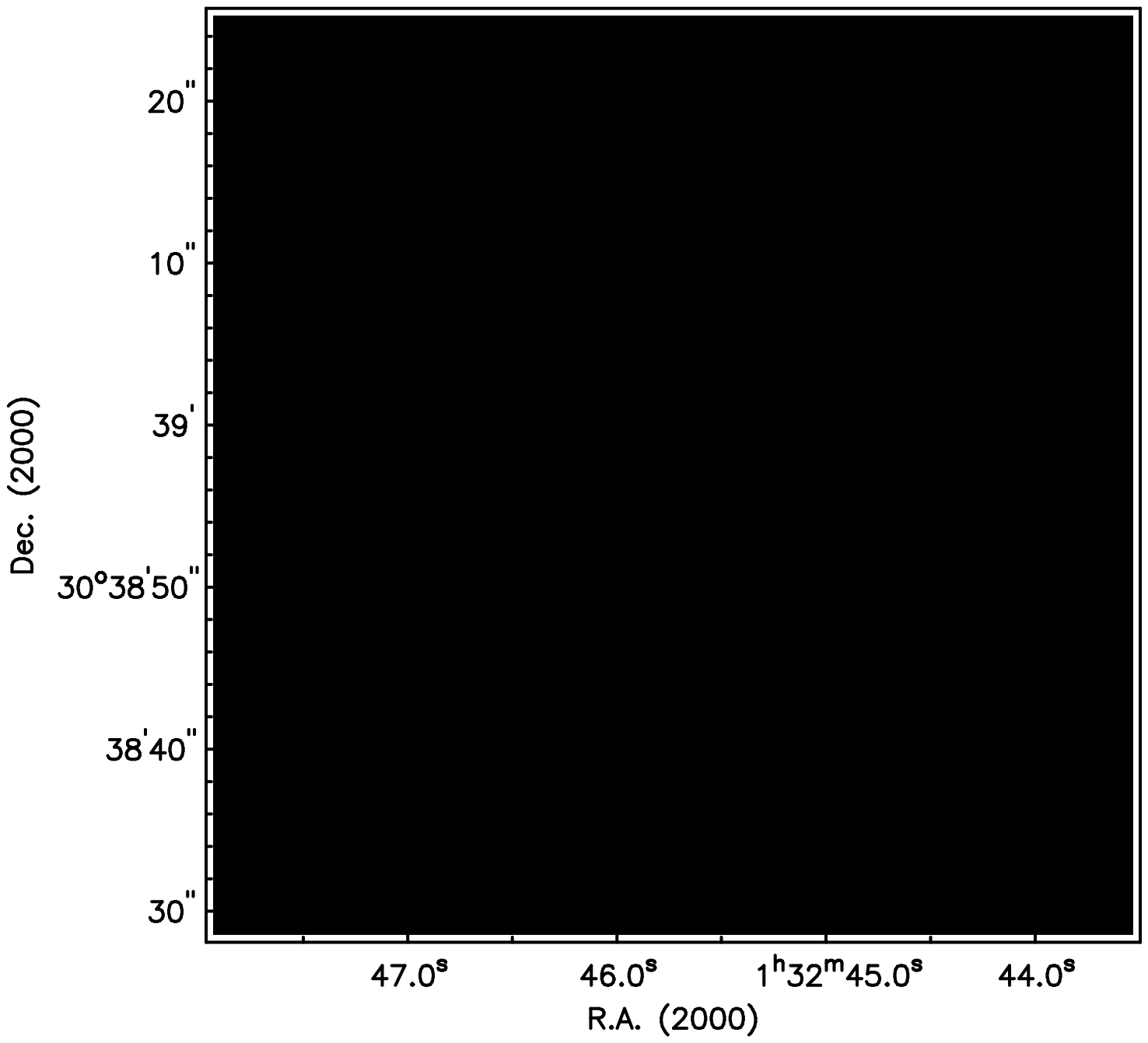}
\includegraphics[width=0.5\textwidth]{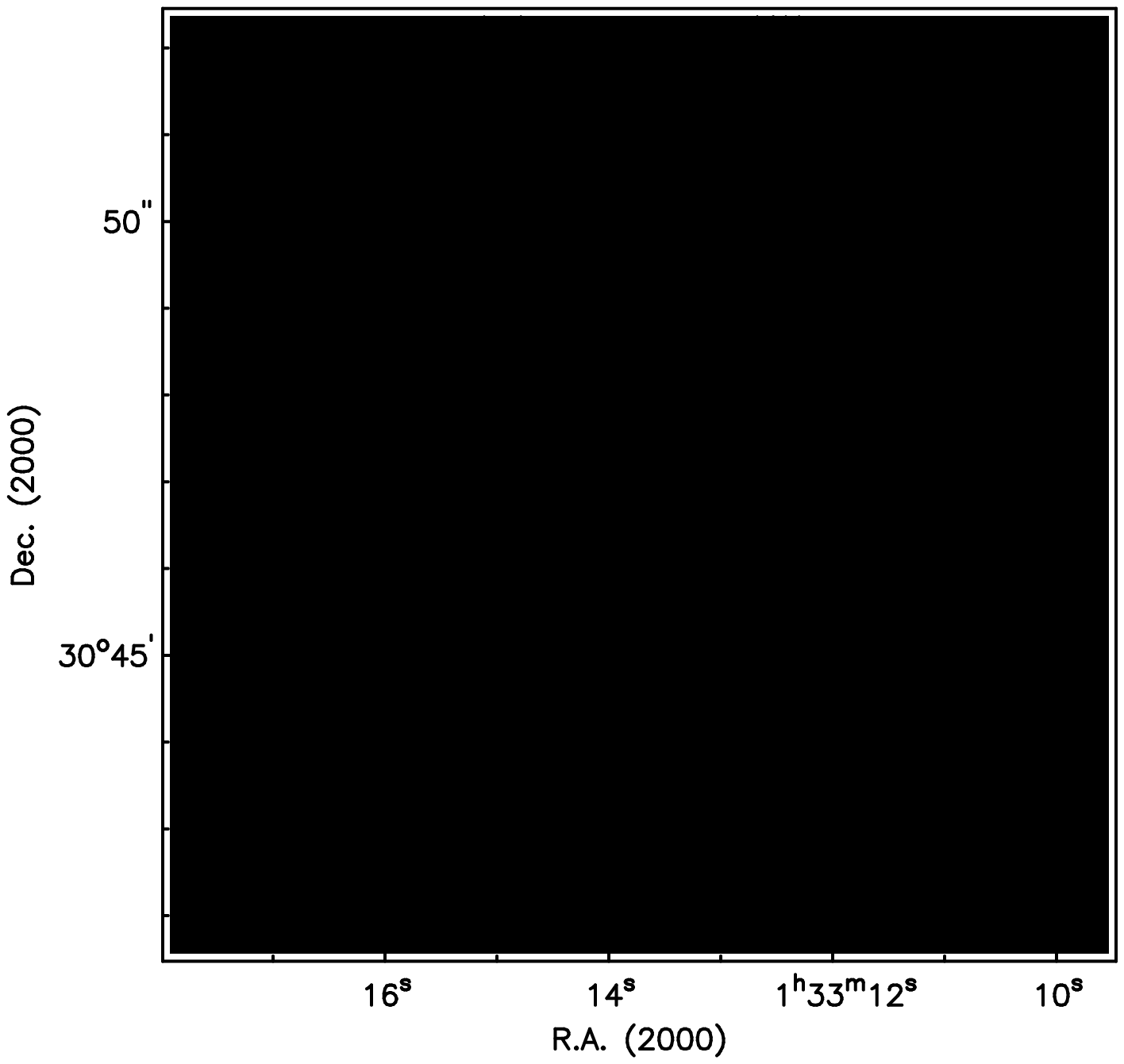} \hfill
\includegraphics[width=0.5\textwidth]{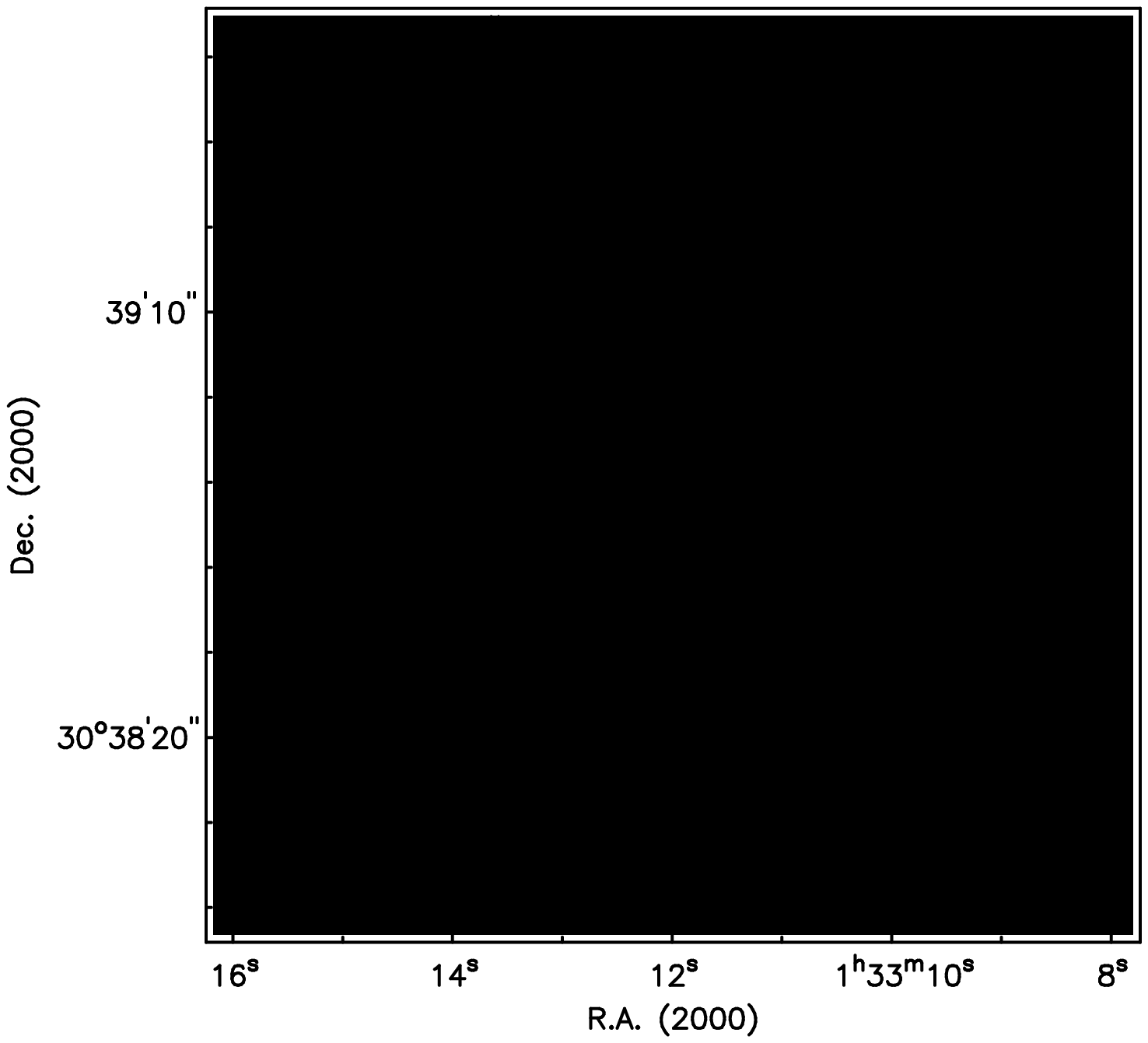}
\caption{Continuum-subtracted \ha\ image of the same set of \hii\ regions as in Figure~\ref{fig3} with FUV emission contours overlaid. The intensity contours are at (5, 10, 20, 40, 60, 80, 95)\% of the maximum intensity, a 5\% level corresponds to 
a range of (6-20)$\sigma$, depending on the region.}
\label{fig4}
\end{figure*}

As one would expect the UV stellar emission follows a different distribution from the dust emission at the 24\,\mi\ and 8\,\mi\ Spitzer bands, and from the emission of the ionized gas at \ha. The UV stellar radiation is generally absorbed by the dust, thus we would not expect to observe UV emission from the stars and dust emission in the same positions within the star forming regions. In Figure~\ref{fig4}, we show the continuum subtracted \ha\ images for our set of \hii\ regions with FUV contours overlaid. The \ha\ emission, in general, surrounds the FUV emission in all \hii\ regions of our sample. This effect is much better appreciated in the shell-like \hii\ regions such as NGC~595, IC 131, NGC~588 and in NGC~604 (Figure~2, upper right panel). The exception to the trend is IC131-West. This knot shows very highly concentrated emission at all wavelengths, it is probably a compact \hii\ region but the extinction derived from different methods (see Table~2) reveals a low dust content in the knot. The existence of UV emission in the center of the shell-like \hii\ regions was also observed by Calzetti et al. (2005) for M51 and Thilker et al. (2005) for M33 in a general study of the UV emission across the galaxy.

The trend observed in all these figures, with the exception of IC131-West, seems to be a {\it stratification} of the different emissions. From inside to the outer border of the region: FUV emission is located at the center, then \ha\ and 24\,\mi\ emissions spatially correlate and both of them are surrounded by the low intensity filamentary 8\,\mi\ emission structure. The best representative cases of the stratification are NGC~595 and NGC~604. For these regions we have also analyzed CO molecular emission and we will comment on the results later.

%-----------------------------------------------------------------------------%
\section{Integrated Flux Measurements}
\label{S4}
%-----------------------------------------------------------------------------%
We have obtained aperture fluxes for our set of \hii\ regions in all the wavelengths available in this study. The apertures used to extract 
the fluxes in each \hii\ region (column 4 in Table~1) were selected to include the total \ha\ emission in our images. Background contamination was eliminated using concentric annuli of 25\arcsec--35\arcsec\ width and internal radii of 25\arcsec--45\arcsec, depending on the aperture size selected for each \hii\ region.
The observed \ha\ luminosities are given in column 9 of Table~1. Based on the noise of the \ha\ image and the aperture size for each \hii\ region we estimated uncertainties in the derived \ha\ luminosities to be in the range of 4\%-12\%. We have checked our photometry with previously reported results provided by Bosch et al. (2002) and Kennicutt (1988) and with the fluxes of the \hii\ region catalogue of M33 (Wyder, Hodge \& Skelton 1997) published in Hodge et al. (2002). We find differences in the \ha\ fluxes up to 3-15\%, which are within the range of the uncertainties of our photometry.

\subsection{Ionization Requirements}

The observed \ha\ luminosity of NGC~604, the most intense \hii\ region in our sample, is given in column 9 of Table~1. It is close to that of 30~Doradus (L(\ha)=5.13$ \times$10$^{39}$ \ergs, Kennicutt  \& Hodge 1986) and corresponds to a stellar content of $\sim$135 equivalent O5(V) stars (using the ionizing photons for an O5(V) star given in Martins et al. 2005). The other \hii\ regions (excepting NGC~595) have \ha\ luminosities one order of magnitude lower than the luminosity of 30~Doradus. Our set of  \hii\ regions are therefore 20-300 times more luminous than the well studied Orion nebula. The \ha\ luminosities of our set of \hii\ regions overlap with the range of luminosities for the \hii\ knots studied by Calzetti et al. (2005) in M51 and by P\'erez-Gonz\'alez et al. (2006) in M81, which allows us to compare our results for individual \hii\ regions with those presented in statistical studies.

The 8\,\mi\ and 24\,\mi\ integrated luminosities for the \hii\ regions are given in Table~1 (columns 7 and 8, respectively). Aperture corrections were applied to the 8\,\mi\ fluxes using the photometric corrections for extended sources given in the SSC web page\footnote{http://ssc.spitzer.caltech.edu/}. For the 24\,\mi\ fluxes the aperture corrections were derived from the theoretical PSF of MIPS at 24\,\mi. The errors in the 8\,\mi\ and 24\,\mi\ fluxes for each \hii\ region are given in Table~1, we estimate flux uncertainties within a range of 3-20\%.

\subsection{Extinction Measurements}

Table~2 is devoted to the global extinction measurements for our set of \hii\ regions, it shows the extinction for each region derived from four different methods: 
A$_{\rm Bal}$(\ha) is the extinction derived using the \ha/\hb\ emission line ratio (Caplan \& Deharveng 1986), A$_{24}$(\ha) is the extinction derived from the ratio between \ha\  and 24\,\mi\ emissions, A$_{\rm rad}$(\ha) is obtained from the ratio between the thermal radio and the \ha\ emission. (Churchwell \& Goss 1999), and A(FUV) is the FUV extinction that will be studied in the next section. 

We have derived A$_{\rm Bal}$(\ha) for NGC~604, using the integrated \ha\ and \hb\ fluxes from the HST images, assuming a temperature of $\rm T_{\rm e}$=8500~K (Esteban et al. 2002), and applying Eq. A.10  in Caplan \& Deharveng (1986) for the case of a screen of homogeneously distributed interstellar dust:  

\begin{equation}
\rm {A_{Bal}(\ha)=5.25 \log (\frac{L(\ha)/L(\hb)}{2.859\times (T_{e}/10^4)^{-0.07}})}
\end{equation}
We obtain A$_{\rm Bal}$(\ha)=0.37$\pm$0.16, which agrees with the values reported by Ma\'iz-Apell\'aniz et al. (2004) (0.24~mag), Viallefond \& Goss (1986) (0.28~mag), and Melnick et al. (1987) (0.39~mag) for the extinctions derived using \ha\ and \hb\ integrated fluxes. In column~2 of Table~2 we give the mean value of the extinctions reported in the literature for our set of \hii\ regions. We take into account only the extinctions derived from integrated \ha\ and \hb\ fluxes. The extinction errors correspond to the standard deviations of the values reported from different authors. For NGC~588: Melnick (1979) obtained 0.46~mag, Viallefond \& Goss (1986) 0.81~mag and Melnick et al. (1987) 0.40~mag; for NGC~592: 0.16~mag and 0.53~mag are reported by Viallefond \& Goss (1986) and Melnick et al. (1987) respectively; for IC131 and IC131-West there are no Balmer extinction values derived from integrated flux measurements, thus we assumed the value given by V\'ilchez et al. (1988) from long-slit observations for these \hii\ regions. For NGC~595 we have assumed a value of 0.27 for the integrated extinction derived from integral-field spectroscopic observations (Rela\~no et al. 2009).

We have been able to calculate A$_{\rm rad}$(\ha) using the radio continuum fluxes at 6 and 20~cm (4.84~GHz and 1.42~GHz) reported by Gordon et al. (1999). These authors estimate the radio spectral index for each \hii\ region: NGC~588, NGC~595 and NGC~604 show a radio spectral index of $\alpha$=0.1 (S~$\propto\nu^{-\alpha}$) and NGC~592, IC131 and IC131-West show spectral indexes of $\alpha$=0.2$\pm 0.1$, $\alpha$=0.2$\pm 0.2$ and $\alpha$=0.2$\pm 0.1$, respectively. Using these values, consistent with previously reported values (Tabatabaei et al. 2007), we assume that our set of \hii\ regions have 
$\alpha\sim$~0.1 and use the 4.84~GHz fluxes in Gordon et al. (1999) to estimate the extinction using the \ha\ luminosities derived in this paper. We have estimated the aperture sizes from the 4.84~GHz image, which is published in Duric et al. (1993), since they are not specified in Gordon et al. (1999). We have then obtained the observed \ha\ fluxes for the same apertures and derived A$_{\rm rad}$(\ha) using Eq.~(2) of Churchwell \& Goss (1999). We assumed a temperature of 10$^4$K for all the \hii\ regions except for NGC~604 that we used $\rm T_{\rm e}$=8500~K. 
The values for A$_{\rm rad}$(\ha) are given in column 4 of Table~2, except for NGC~588 for which we are not able to obtain an accurate value for the extinction. The errors quoted for the extinctions are a combination of the photometric errors and the extinction uncertainties for a change of 20\% in the estimated aperture radii. 
For NGC~592 and NGC~604 we find values within the range reported in the literature (Israel \& Kennicutt 1980, Viallefond \& Goss 1986 and Churchwell \& Goss 1999). For IC131-West we find an extinction 0.5~mag lower than the value given by Viallefond \& Goss (1986), this is due to the different aperture of the \ha\ and radio fluxes used by Viallefond \& Goss (1986) to derive the extinction for this region. Taking into account the uncertainties, the extinction given for NGC~595 is close (a difference of 0.14~mag) to the value reported by Viallefond et al. (1983) for a similar aperture.

A$_{24}$(\ha) is the extinction derived assuming that the absorbed \ha\ luminosity scales with the 24\,\mi\ luminosity (Kennicutt et al. 2007). In this case the corrected \ha\ luminosity will be a combination of the observed \ha\ luminosity and the 24\,\mi\ luminosity, and the extinction can be derived in the usual way: 

\begin{equation}
\rm {A_{24}(\ha)=-2.5 \log (\frac{L_{obs}(\ha)}{L_{obs}(\ha)+a\times L(24\,\mi)})}
\end{equation}
where a=(0.031$\pm$0.006) is a scale factor that is empirically obtained from flux measurements of HII knots in different galaxies (Calzetti et al. 2007). We have obtained A$_{24}$(\ha) using the luminosities given in Table~1. With the exception of NGC~604 and NGC~592, the extinctions derived using the 24\,\mi\ and \ha\ emissions and those derived from the radio continuum and \ha\ fluxes agree within the uncertainties. For NGC~604 and NGC~592, increasing the \hii\ region temperature by 1500~K would decrease A$_{\rm rad}$(\ha) by 0.1~mag, reducing the discrepancy between A$_{\rm rad}$(\ha) and A$_{24}$(\ha).

%-----------------------------------------------------------------------------%
\subsection{Stellar Masses}
\label{S4.1}
%-----------------------------------------------------------------------------%
The excellent spatial correlation between the 24\,\mi\ and the absorbed \ha\ emission in NGC~604 (see Figure~2 lower right panel), which was derived using the Balmer extinction map shown in Figure~10 supports the hypothesis that the 24\,\mi\  emission is a good tracer for the extincted star formation (Calzetti et al. 2007; P\'erez-Gonz\'alez et al. 2006). The linear relation between these two luminosities implies that 
the dust emitting at 24\,\mi\ is dominated by the light arising from the star clusters that also produce most of the ionizing luminosity. Here, we show that even at small spatial scales the 24\,\mi\ luminosity traces the current star formation that ionizes the interstellar gas. 
The correlation is not as good for 8\,\mi\ and the UV; 8\,\mi\ emission has a more extended 
filamentary structure than \ha\ emission and UV is more concentrated in the interior of the \ha\ 
shells where there is no apparent ionized gas (see Figures~3 and 4). 

\begin{figure*}
\includegraphics[]{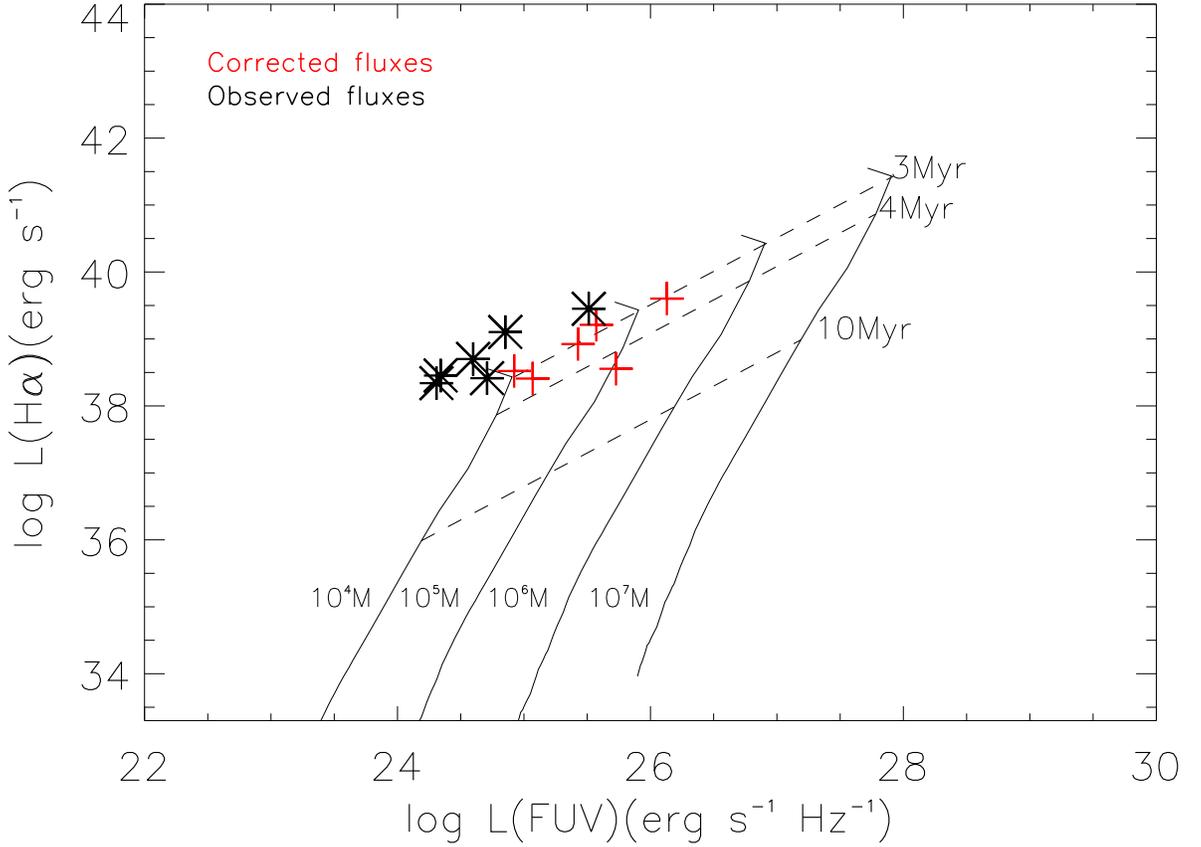}
\caption{Temporal evolution of the logarithmic \ha\ luminosity versus logarithmic FUV luminosity for a set of instantaneous star formation burst of different stellar masses (full lines). The models were obtained with Starburst99 (Leitherer et al. 1999) using a Salpeter IMF with mass limit 0.1-100$\msun$ and metallicity Z=0.02. The dashed lines show the luminosities at 3, 4 and 10\,Myr after the start of the burst. Black points are observed luminosities and red points are the extinction--corrected luminosities for the set of \hii\ regions. The models describe the ranges of masses and ages for our sets of \hii\ regions; for NGC~604, the most luminous \hii\ region, we find agreement with previously reported ages (e.g. $\tau\sim$3\,Myr for NGC~604,  Gonz\'alez-Delgado \& P\'erez 2000)}
\label{fig5}
\end{figure*}

In order to analyze more deeply these observed trends, we have quantified the star formation in each individual \hii\ region using FUV, 24\,\mi\ and  8\,\mi\ emissions, and compared the results with those derived from the extinction--corrected \ha\ luminosities. We will then be able to compare our predictions of SF measurements for \hii\ regions with the results derived from statistical studies applied to regions in different galaxies. The SFR calibrations widely used to quantify the SF are derived assuming a constant SFR over a time scale of 100\,Myr (Kennicutt 1998, Calzetti et al. 2007, Iglesias-P\'aramo et al. 2006), but the typical ages for the \hii\ regions are much smaller ($\sim$2-8\,Myr, Bresolin \& Kennicutt 1997, Copetti et al. 1985) and the observable parameters suggest in general an instantaneous burst of star formation (e.g. Gonz\'alez-Delgado \& P\'erez 2000, Malamuth et al. 1996). These properties are also shown in Figure~\ref{fig5}, where we compare the extinction--corrected \ha\ and FUV luminosities derived for our \hii\ regions with a set of instantaneous SF bursts of different stellar masses obtained using Starburst99 (Leitherer et al. 1999). The extinction--corrected luminosities correspond to models of \hii\ regions having stellar masses $\sim$10$^4$-10$^5$\msun\ and ages of 3-6\,Myr. Although we cannot rule out the possibility that \hii\ regions might have non-coeval star clusters (e.g. 30 Doradus among other \hii\ regions shows evidence of a new stellar generation, see Brandner et al. 2001), we assume as approximation an instantaneous burst of SF to describe our \hii\ regions, which furthermore allows us to define their ages. In the instantaneous SF burst approximation the SFR is
infinite at the beginning of the burst (t$=$0), and remains zero after this time.
In order to study the amount of SF within the \hii\ regions, the classical calibrations of the SFR valid for galaxies with continuous SF (Kennicutt 1998) are not applicable here, as we will explain further in this section. Thus, we will derive new calibrations that allow us to quantify the SF from the extinction--corrected \ha\ and FUV luminosities. 

With the exception of IC131, there are independent estimates of the age and the IMF exponent for the \hii\ regions of our sample (e.g. Hunter et al 1996, Drissen et al. 1993, Terlevich et al. 1996, Gonz\'alez-Delgado \& P\'erez 2000 for NGC~604; Malamuth et al. 1996 for NGC~595; Jamet et al. 2004 for NGC~588 and Pellerin (2006) for NGC~592). Based on these estimates we can assume that our \hii\ regions are $\tau\sim$4\,Myr old and have Salpeter IMFs ($\alpha$=2.35) with mass limits 0.1-100$\msun$. From our Figure~\ref{fig5} it is clear that our \hii\ regions span an age range of 3-6\,Myr; the assumed age of 4\,Myr is an approximation that is supported by the ages estimated in the literature for some of our \hii\ regions: Gonz\'alez-Delgado \& P\'erez (2000) predicts an age of 3\,Myr for NGC~604 using photoionization models, Malamuth et al. (1996) gives an age of 4.5\,Myr for NGC~595 and Jamet et al. (2004) an age of 4.2\,Myr for NGC~588, both studies used color-magnitude diagrams to predict the ages. Pellerin (2006) predicts an age of 4\,Myr for NGC~592 based on FUV spectral synthesis analysis and finds consistent ages for the rest of the \hii\ regions. There are no estimated ages for IC131 and IC131-West in the literature but from Figure~\ref{fig5} we can assume that 4\,Myr could also be a reasonable estimated age for them.

We have then used Starburst99  to derive the star formation calibrations for our \hii\ regions in the instantaneous burst approximation using models with Salpeter IMF, mass limits of 0.1-100$\msun$ and metallicity Z=0.02. At $\tau\sim$4\,Myr the relation between the total stellar mass and the \ha\ luminosity will give: 
\begin{equation}
\rm {SF(\ha)(\msun)=1.29\times10^{-34}L(\ha)(\ergs)}
\end{equation}
and for the luminosity at $\lambda$1516\AA\ we obtain the following relation:
 
\begin{equation}
\rm {SF(FUV)(\msun)=1.64\times10^{-21}L_{FUV}(erg s^{-1}Hz^{-1})}
\end{equation}
We have applied these calibrations to obtain the SFs for our \hii\ regions using the set of wavelengths we are considering in this paper. The  observed \ha\ luminosities were corrected for extinction using a mean value of A$_{\rm Bal}$(\ha) derived from the literature (listed in column~2 of Table~2). This method gives uncertainties in the corrected  \ha\ luminosities of $\sim$25\%. For the case of NGC~604, we use the value derived here, A$_{\rm Bal}$(\ha)=0.37 to obtain the extinction--corrected \ha\  luminosity. The extinction--corrected \ha\ luminosities are given in the last column of Table~1 and the SFs derived from them are given in column 4 of Table~3.   A further uncertainty in quantifying the SF is the possible error in the age of the \hii\ region. We 
estimate the error to be $\sim$20\% for the FUV calibration and $\sim$60\% for the \ha\ calibration in our case, because the ages of the \hii\ regions have been estimated by different methods to lie between 
3 and 4\,Myr.  

We use the empirical correlations between the  24\,\mi\ and extinction--corrected  \ha\ luminosity 
found by Calzetti et al. (2007) for their high-metallicity data points -the metallicity for 
most of our \hii\ regions are in the higher range defined by these authors (see Magrini et al. 2007; V\'ilchez et al. 1988)- to derive 
the SF from the 24\,\mi\ luminosity, SF(24\,\mi). The SFs derived in this way are given in column 3 of Table~3. These are lower by a factor of 0.1-0.6 (see column 7 in Table~3) than SF(\ha$^{\rm corr}$). This result supports the suggestion of P\'erez-Gonz\'alez et al. (2006) that the 24\,\mi\ emission is not tracing the total SF but only traces the ionizing photons absorbed by dust, a suggestion that is based on the better correlation they find between the 24\,\mi\ and the absorbed (extincted) \ha\ luminosity than to the extinction--corrected \ha\ luminosity (see Figure~8 in P\'erez-Gonz\'alez et al. 2006) for the \hii\ emission knots in M81. 

Following these ideas, Kennicutt et al. (2007) proposed a combination of \ha\ and 24\,\mi\ luminosities as a better tracer of the total SFR. They proposed that the observed \ha\ luminosity traces the unobscured star formation, while the 24\,\mi\ emission represents the star formation reprocessed by dust. Thus, the extinction--corrected \ha\ luminosity is then expressed as a linear combination of these two luminosities: L$_{\rm corr}$(\ha)=L$_{\rm obs}$(\ha)+a$\times$L(24\,\mi). We have used a=(0.031$\pm$0.006) (Calzetti et al. 2007) to obtain the linear combination of both luminosities and to derive the total SF, SF(com). The results are given in column 5 of Table 3 and the comparison with the SF(\ha$^{\rm corr}$) is given in column 8 of this table. The star formation derived from the {\it combined} luminosities represents 60\%-100\%\ of the SF(\ha$^{\rm corr}$). For NGC~588, SF(com) represents only 63\% of the SF(\ha$^{\rm corr}$), for this \hii\ region we have derived a value of A$_{24}$(\ha) which is 0.5~mag lower than the Balmer extinction derived from the literature (see Table~2).

Calzetti et al. (2005) and P\'erez-Gonz\'alez et al. (2006) did not find as good correlation between the integrated 8\,\mi\ and 
\ha\ luminosities for the \hii\ regions in M51 and M81 as they found for 24\,\mi\  and \ha\ luminosities. 
They suggested some mechanisms that can affect the relation between the 8\,\mi\ and \ha\ luminosities (contamination of 
diffuse emission from the general galactic radiation field and/or destruction of the PAH emitters at 8\,\mi\ in environments of high intensity radiation fields). We have derived the SFs for 
our \hii\ regions using the  8\,\mi\ luminosity and the relation found by Calzetti et al. (2005) (Eq.~(7) in that paper). The result is  given in column 2 of Table~3. The SFs derived from the 8\,\mi\ emission are clearly much lower than the SF(\ha$^{\rm corr}$). 

Ultraviolet emission is strongly affected by dust attenuation. In order to use it as a SF tracer, the UV emission has to be corrected using a reliable estimation of the extinction in this wavelength range. Generally, the attenuation has been estimated from the slope of the spectrum in the UV, $\beta$, which correlates well with the ratio between the IR and UV emissions for starburst galaxies (Meurer et al. 1999), but shows significant dispersion for normal galaxies (Kong et al. 2004; Cortese et al. 2006; Buat et al. 2005). We have estimated here the extinction in the UV wavelength range using Starburst99 models in the instantaneous burst approximation to compute the intrinsic (FUV-NUV) color and comparing it with the observed (FUV-NUV) color for our \hii\ regions. Assuming a given age for the starburst, the difference of the observed and intrinsic (FUV-NUV) color allows us to estimate the color excess E(B-V) following the starburst reddening curve given in Eq.~(4) of Calzetti et al. (2000), and then to derive the FUV extinctions. 
The intrinsic (FUV-NUV) color varies slightly with age: for 2~$<\tau<$~4\,Myr the standard deviation of the intrinsic (FUV-NUV) color derived from our models is $\sigma$=0.06 and for 4~$<\tau<$~20\,Myr,  $\sigma$=0.03. Assuming an age of $\tau$=4\,Myr, we have derived from our models an intrinsic (FUV-NUV) color of -0.101 and then, using the observed (FUV-NUV) color for each \hii\ region, we have obtained the FUV extinctions. The results are listed in column~5 of Table~2. The errors quoted in the table take into account the photometric errors and the uncertainties in the zero-point calibration,  $\pm$0.05~m$_{AB}$ and $\pm$0.03~m$_{AB}$ for FUV and NUV respectively (Morrissey et al. 2007). As a check, we have compared the color excess obtained from this method with the color excess derived from other techniques given for three \hii\ regions (NGC~588, NGC~595 and NGC~604) in the literature, we found good agreement within the uncertainties in all these regions. We also derived the extinction at \ha, assuming the same color excess E(B$-$V) for the nebular gas as for the stellar continuum, and we found good agreement within the uncertainties with the extinctions at \ha\ listed in column~2 of Table~2. We expect an additional uncertainty in the derived FUV extinctions related to the extinction curve used here (see Calzetti (2001) for a comparison of different extinction curves). Assuming extinction curves for the 30~Doradus region in the Large Magellanic Cloud (Fitzpatrick 1985) and the Small Magellanic Cloud's bar (Gordon \& Clayton 1998), we expect deviations of the FUV extinctions reported here of $\sim$10\% and $\sim$50\%, respectively.

We have used the FUV extinctions to correct the observed FUV luminosities and then we have quantified the SF (SF(FUV$^{\rm corr}$)) for each \hii\ region using Eq.~(4). The result is listed in column~6 of Table 3. We find consistent values within the uncertainties given here between the SF(\ha$^{\rm corr}$) (the SF derived from the observed \ha\ luminosity corrected for extinction using the Balmer extinction values derived from the literature (column~2 of Table~2)) and SF(FUV$^{\rm corr}$) (see last column in Table~3). This shows that the SF(\ha$^{\rm corr}$) and SF(FUV$^{\rm corr}$) agree well within the uncertainties assuming an age of 4~Myr for the \hii\ regions, which is quite reasonable given the ages reported for these \hii\ regions in the literature (see above). Applying the same method for an age of 3~Myr, which gives an intrinsic (FUV-NUV) color in our models of -0.119, we obtain values for the SF(\ha$^{\rm corr}$)/SF(FUV$^{\rm corr}$) between 0.6 and 1.2 (except for NGC~595 where we obtained a value of 0.2).

We report here the SFs for our set of \hii\ regions using wavelengths from the UV to the IR and assuming the \hii\ regions are formed in an instantaneous burst of star formation and are $\sim$4\,Myr old. 
The SF derived using the 8\,\mi\ emission is much lower than the SF predicted using the extinction 
corrected \ha\ luminosity. This result is not surprising because the 8\,\mi\ emission is indeed only poorly correlated with the \ha\ emission at the small scales that we are exploring here in our set of \hii\ regions (see Figure~3). The 24\,\mi\ emission also gives lower values of the SFs than the SF(\ha$^{\rm corr}$), which gives evidence of the low dust content within these \hii\ regions suggested by the low extinctions derived for them (see Table~2).
The combination of 24\,\mi\ and observed \ha\ luminosities gives values of the SFs close to those derived from the extinction--corrected \ha\ luminosity. This result, which is valid for our set of \hii\ regions, agrees with the results reported for other galaxies (Calzetti et al. 2007, Zhu et al. 2008) and is supported by the very good spatial correlation between the \ha\ and the 24\,\mi\ emission observed in our \hii\ regions (see Figure~1). Given the uncertainties in the SF derived from the UV emission we also find good agreement between SF(FUV$^{\rm corr}$) and SF(\ha$^{\rm corr}$) (see last column in Table~3). 

The application of the classical SFR calibrations valid for galaxies with continuous star formation over time scales of $\sim$100\,Myr (Kennicutt 1998) gives however different results than those shown here. The FUV emission overpredicts the SFRs derived from the extinction--corrected \ha\ luminosities in the continuous SF approximation. This effect was also observed by Sullivan et al. (2000), who explained the difference by a series of starbursts superimposed on the galactic star formation history (see also Bell \& Kennicutt 2001). We show here that the assumption of an instantaneous burst of SF is a better approximation in quantifying the SF for an \hii\ region than the assumption implied in the classical SFR calibrations. At 100\,Myr the \ha/FUV ratio drops by a factor of $\sim$2 (Starburst99, Leitherer et al. 1999), because at this time the ratio is dominated by the FUV emission coming from an older stellar population. Thus, the FUV emission will in general overestimate the actual SFR. As a test we have obtained Starburst99 models assuming a constant SFR and the same IMF as our previous models and applied the same method to derive the SFRs using the extinction--corrected \ha\ and FUV luminosities. We find consistent values (SFR(\ha$^{\rm corr}$)/SFR(FUV$^{\rm corr}$)$\sim$1) at much later times ($\tau\ga 15$\,Myr), which implies unphysical ages for the \hii\ regions. These results show that caution must be taken when applying the classical SFR calibrations to environments where the continuous SF approximation is not applicable, as in  the case of the \hii\ regions. 

%-----------------------------------------------------------------------------%
\section{NGC~604 and NGC~595}
\label{S5}
%-----------------------------------------------------------------------------%

NGC~604 is the most luminous and most studied \hii\ region in M33. It shows a 
very complex \ha\ morphology of shells and filaments revealing a complicated 
kinematic environment that has been studied by several authors (e.g. 
Rosa \& Solf 1984; Sabalisck et al. 1996; Yang et al. 1996; Medina-Tanco 
et al. 1997). Tenorio-Tagle et al. (2000) modelled the big cavities and show 
evidence of shells blowing out into the halo of M33. Stellar photometry of the region 
reveals a young population of 3-5\,Myr (Hunter at al. 1996) and evidence of 
WR-stars has been found by several authors (Hunter et al. 1996; Drissen et al. 1993; 
D'Odorico \& Rosa 1981; Rosa \& D'Odorico 1982). Using radio observations and comparing them to the \ha\ emission,  
Churchwell \& Goss (1999) obtained an optical depth map of NGC~604. The visual extinction varies across the 
face of the region, with the maxima located at the position of intense radio knots. This suggests
the presence of dust embedded within the hot ionized gas in each radio component.
The spatial distribution of ionized gas and dust using optical data has been exhaustively studied by
Ma\' iz-Apell\' aniz et al. (2004) revealing differences between the extinction map derived from 
radio and \ha\ observations and the extinction map obtained from the Balmer decrement. 
Wilson \& Scoville (1992) studied the distribution of the molecular gas in NGC~604 and identified 
four molecular clouds in the region, some of them showing a high CO(J=3-2)/CO(J=1-0) ratio that would 
correspond to high temperature and density conditions of compressed gas where new stars 
could be forming (Tosaki et al. 2007).

\begin{figure*}
\includegraphics[width=0.5\textwidth]{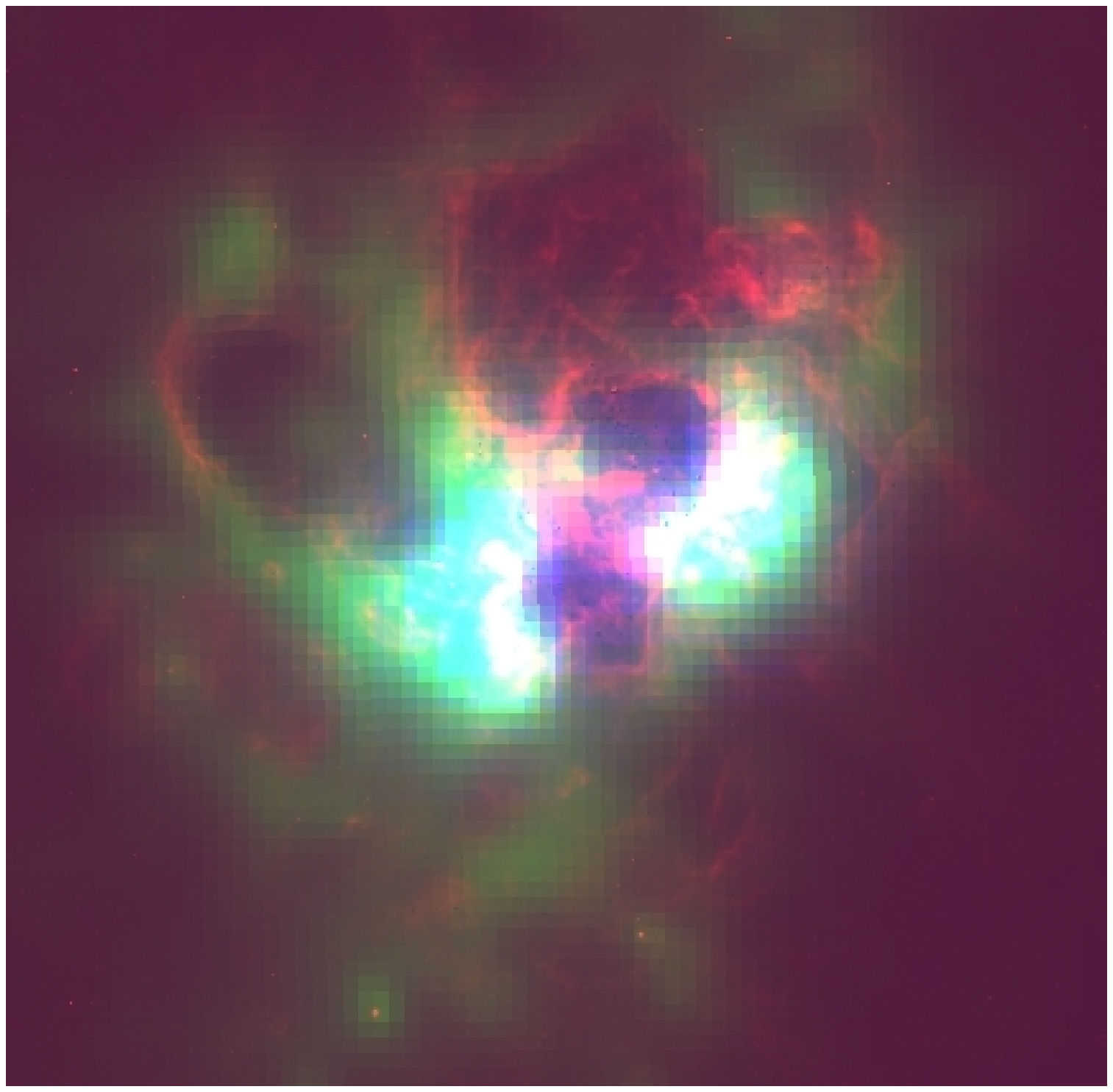}
\includegraphics[width=0.5\textwidth]{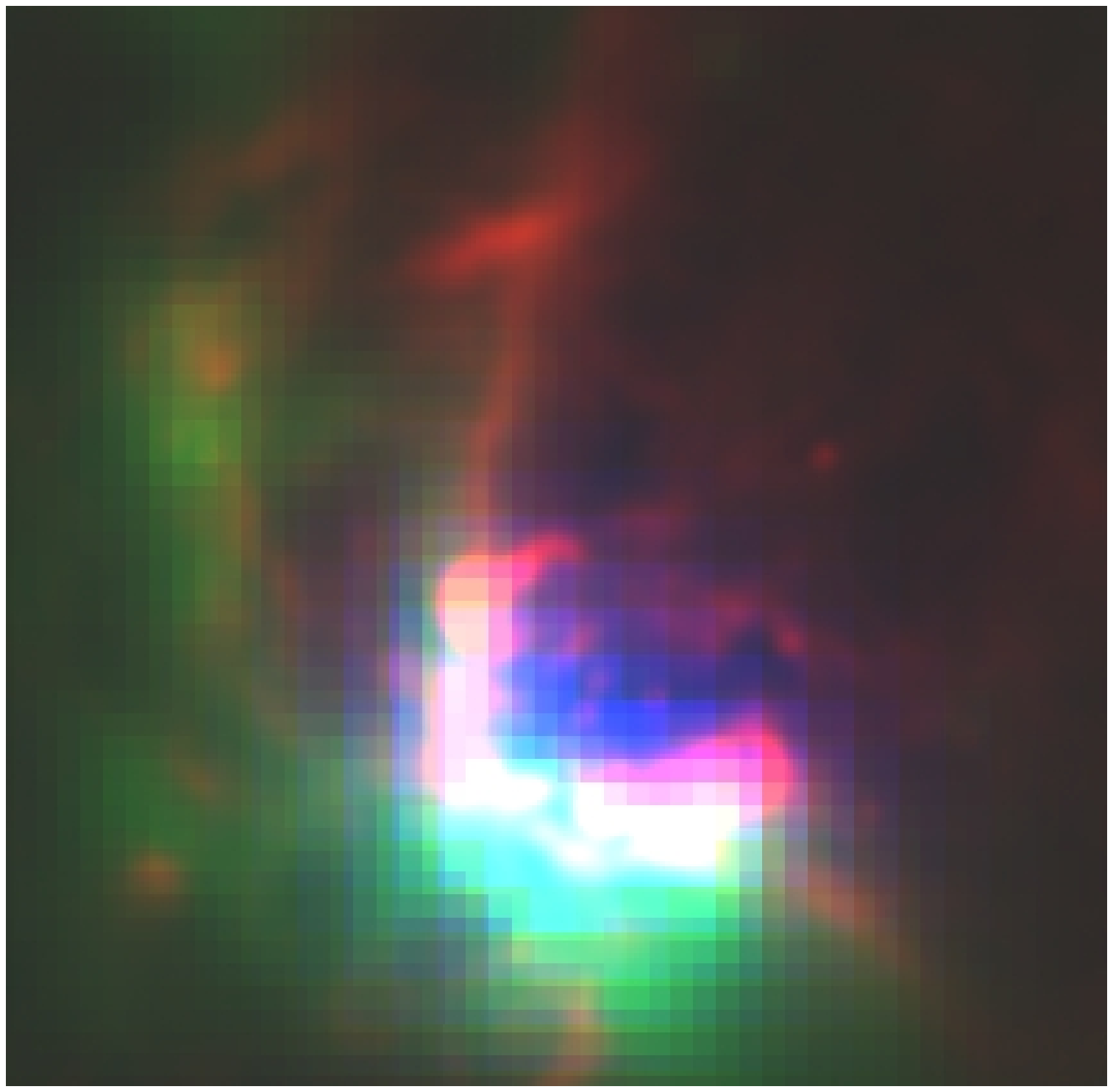}
\caption{Three color image (\ha\ (red), 8\,\mi\ (green), 24\,\mi\ (blue)) of NGC~604 (left) and NGC~595 (right). While the 8\,\mi\ emission delineates the \ha\ shells and filaments across the face of the \hii\ regions, the 24\,\mi\ emission is more compact and corresponds spatially with the most intense \ha\ knots inside the regions.}
\label{fig5b}
\end{figure*}

The Spitzer observations of NGC~604 presented in this paper and the comparison with observations presented in previous studies allow us to perform a deeper study of NGC~604. We are able to compare the extinction maps derived from different methods with the dust emission within the \hii\ region, the location of the molecular gas and the position of the stars producing the UV and \ha\ emissions. The comparison will give a more  complete picture of the structure of the \hii\ region and will allow us to suggest places where a new generation of stars could be located. In Figure~\ref{fig5b} (left) we show a color illustration of the emission structure of NGC~604: the arcs and filaments in red corresponding to \ha\ emission are surrounded by the 8\,\mi\ emission in green. The 24\,\mi\ emission is restricted to the central part of the region and correlates with the position of the most intense central \ha\ knots. NGC~595, the second most luminous \hii\ region in M33 after NGC~604, is less well studied. We show in Figure~\ref{fig5b} (right) a three-color image of the region. The emission distribution at \ha\ (red), 8\,\mi\ (green) and 24\,\mi\ (blue) is similar to that found for NGC~604. Unfortunately, we are not able to derive extinction maps for NGC~ 595 as for NGC~604, but we have CO molecular data available for this region from Wilson \& Scoville (1992) that will be compared with observations at other wavelengths.

\subsection{Comparison to CO Emission}

\begin{figure*}
\includegraphics[width=0.55\textwidth]{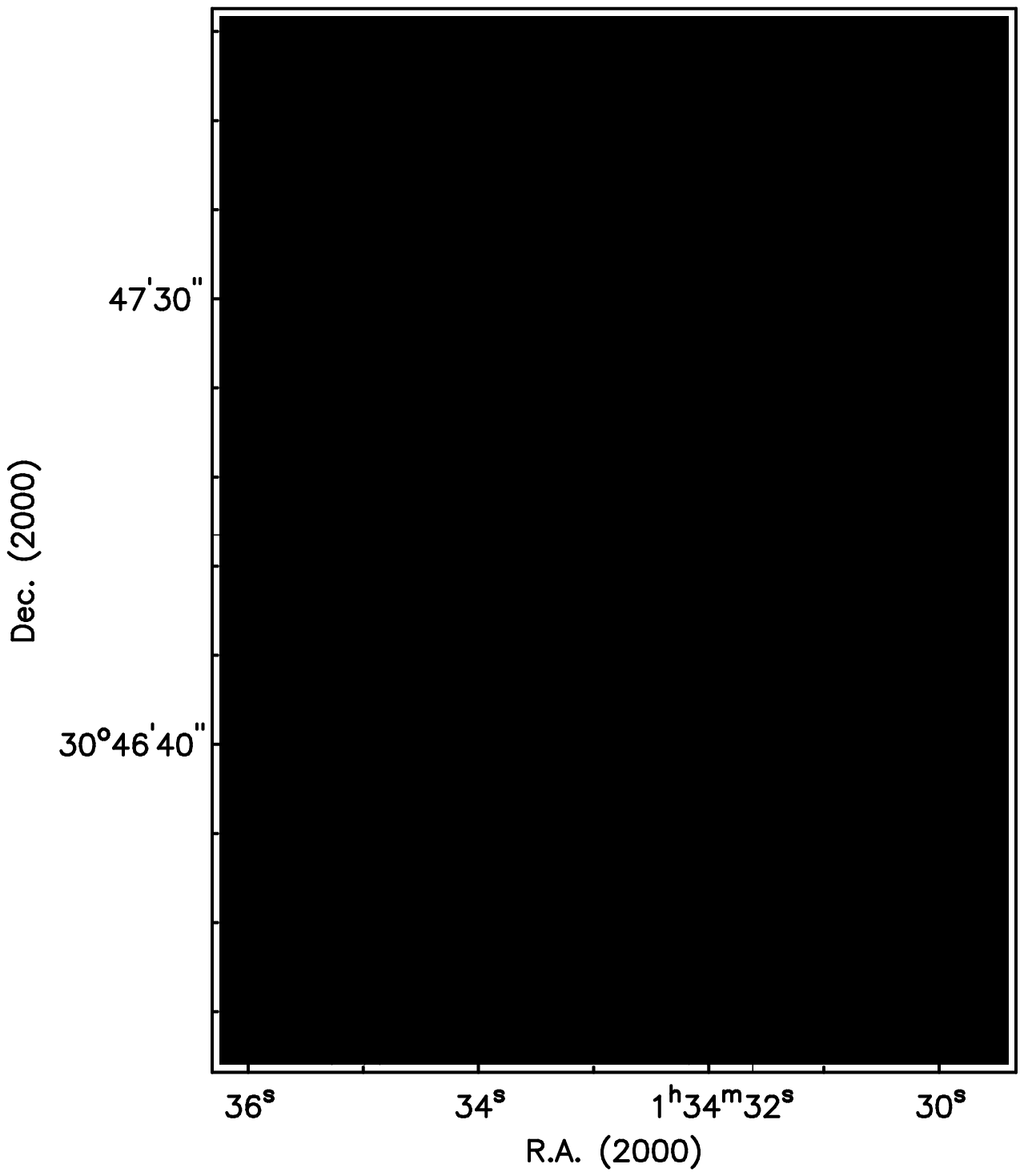}
\includegraphics[width=0.55\textwidth]{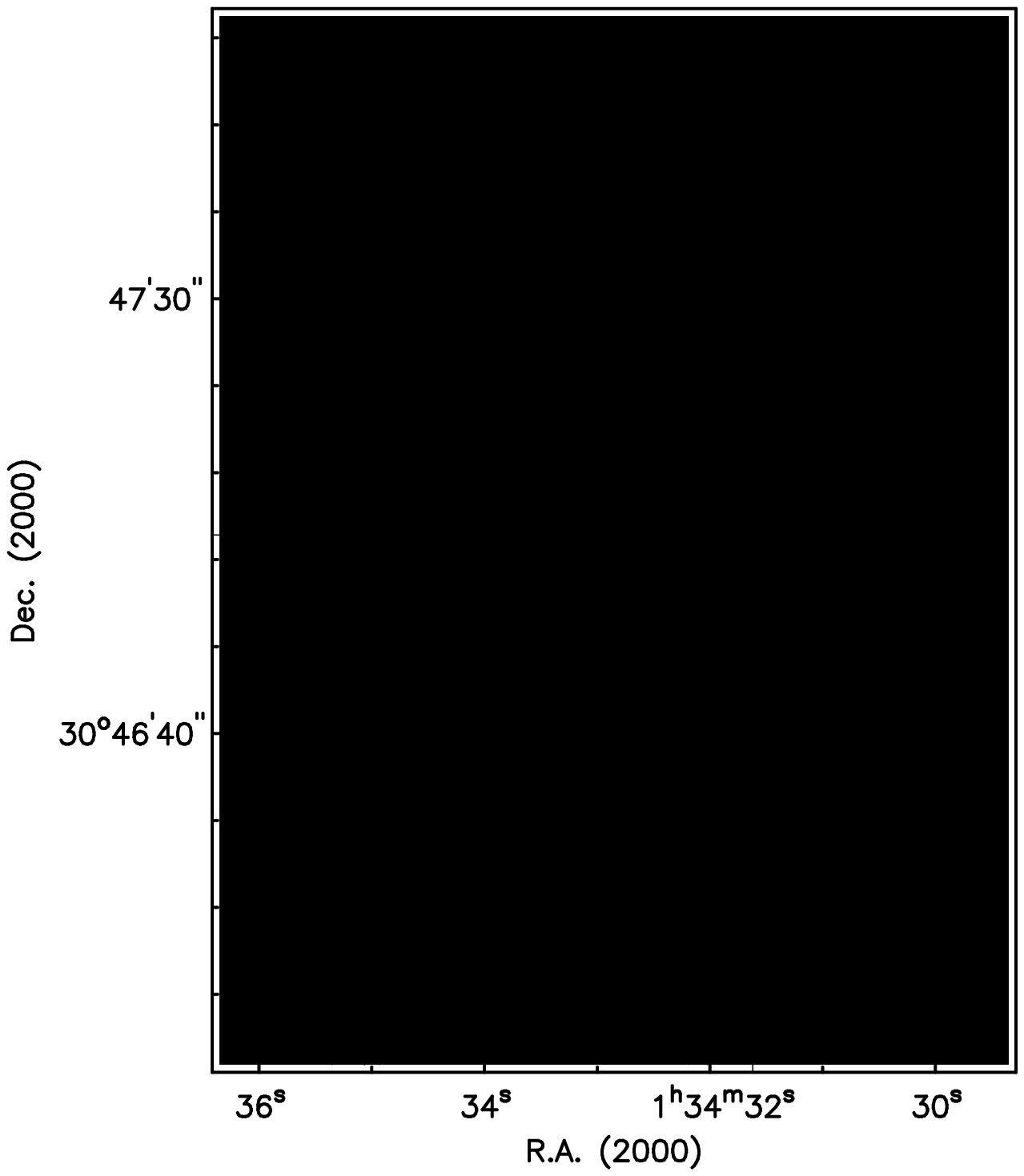}
\includegraphics[width=0.55\textwidth]{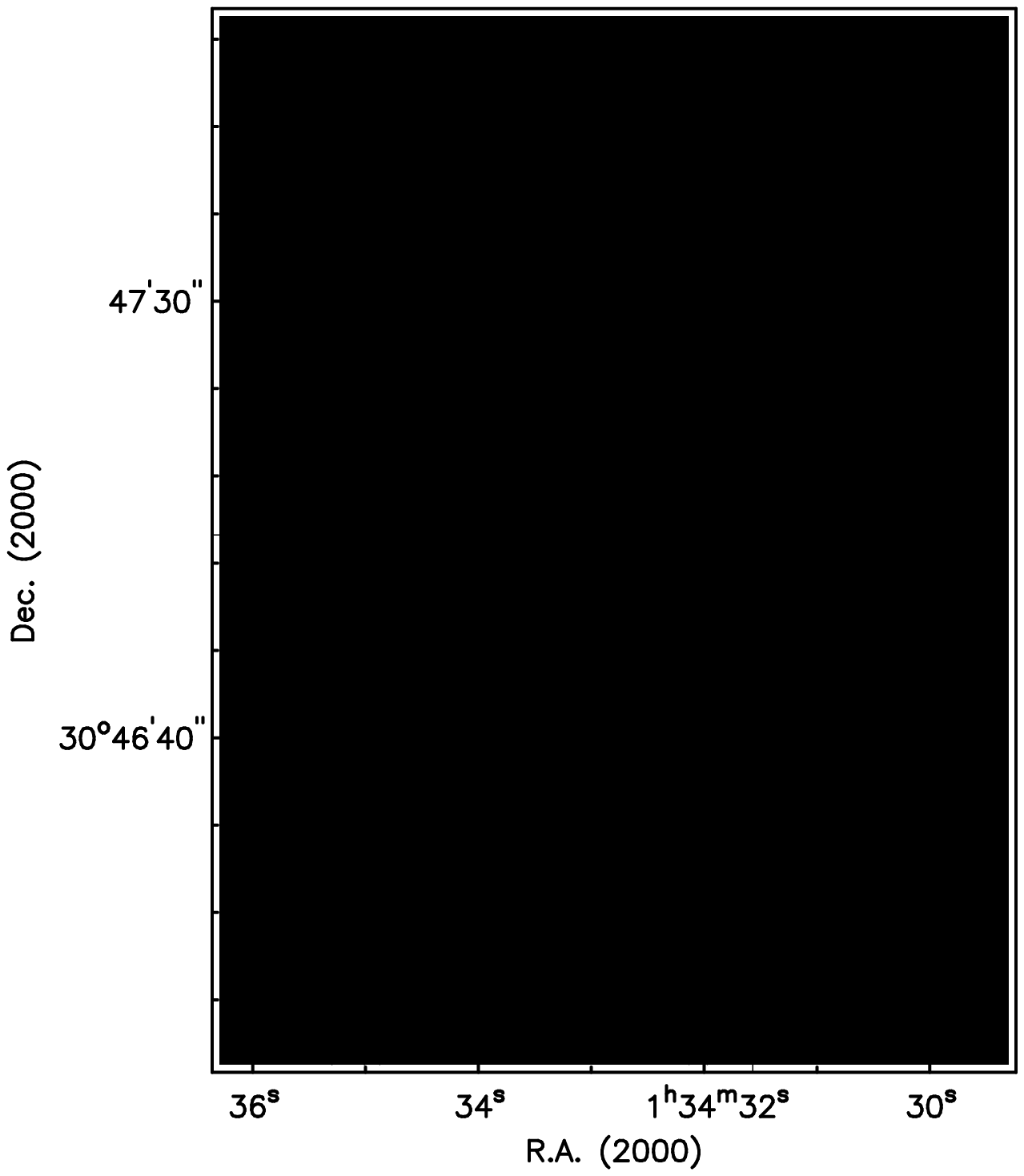} 
\includegraphics[width=0.55\textwidth]{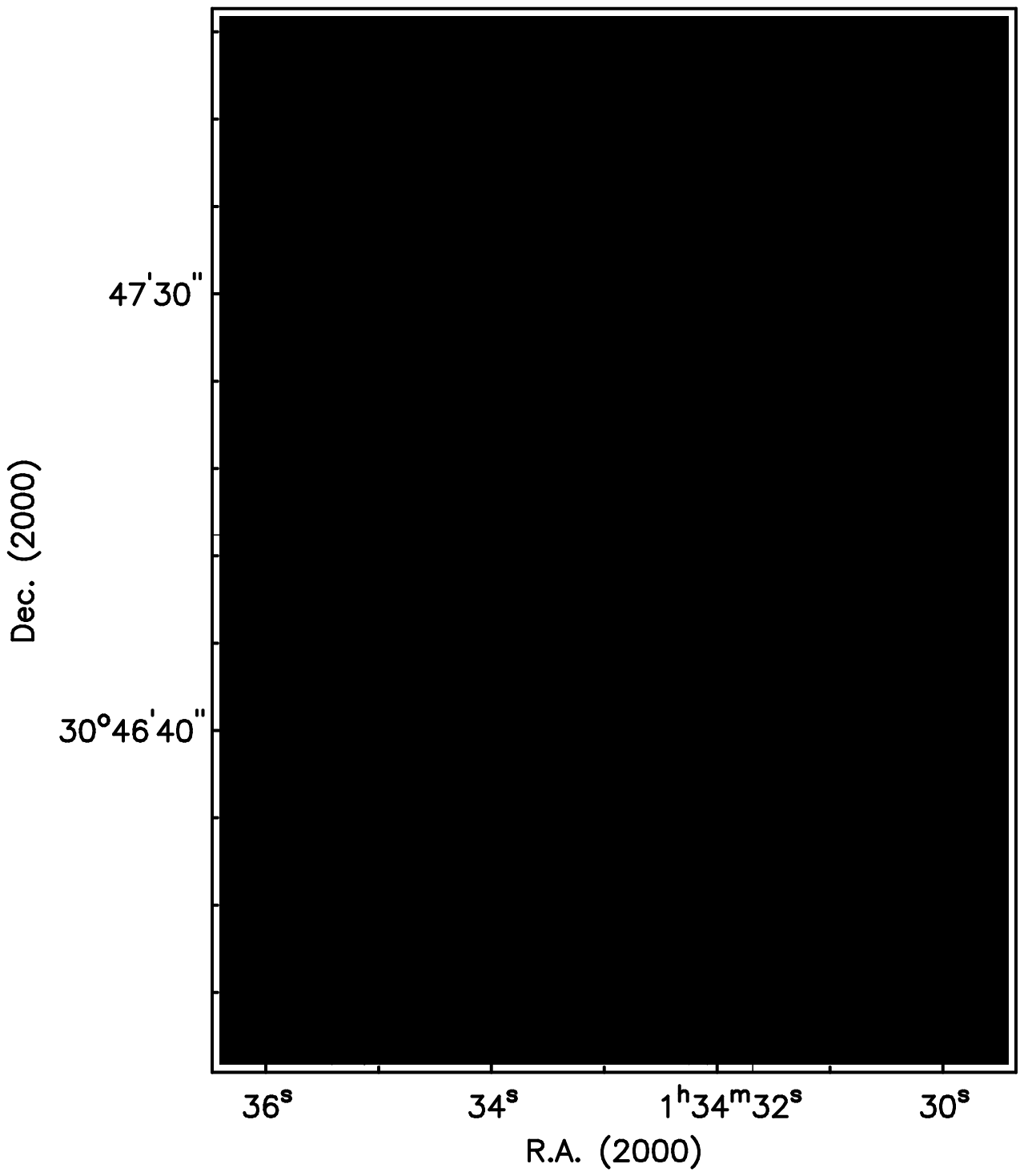}
\caption{Comparison of CO observations with IR, \ha\  and FUV emission distributions for NGC~604. In color we show the 24\,\mi\ (top-left), 8\,\mi\ (bottom-left), \ha\ (top-right) and FUV (bottom-right) emission distributions. Contours correspond to the CO observations of Wilson \& Scoville (1992). The capital letters correspond to the position of radio knots identified in Churchwell \& Goss (1999). The CO observations only cover the south-east part of the \hii\ region delineated by the black axis. We do not have information of the CO emission distribution for the northern part of the \hii\ region.}
\label{fig6}
\end{figure*}

For NGC~604 we compare the location of the radio knots identified by Churchwell \& Goss (1999) and the CO emission distribution from Wilson \& Scoville (1992) with emission at 24\,\mi\ and 8\,\mi\ in the mid-infrared, FUV emission and ionized gas emitting at \ha\ (see Figure~\ref{fig6}). The spatial distribution at these wavelengths shows a similar behavior as in the rest of the \hii\ regions in our sample (see also Figure~\ref{fig2}). The FUV emission, located within the larger \ha\ shells and probably tracing a moderate age stellar population, anticorrelates with the CO molecular emission (lower right panel in Figure~\ref{fig6}). Except for the radio knot E, the faintest one, the radio components coincide with the location of maxima at 24\,\mi\ emission and have \ha\ emission associated with them (see top panels in Figure~\ref{fig6}). The eastern high intensity knots at 24\,\mi\ and 8\,\mi\ coincide with the position of the most intense central molecular cloud, but the maximum at 8\,\mi\ emission seems to be closer to the maximum of the molecular cloud than the maximum at 24\,\mi\ (upper and lower panel in Figure~\ref{fig6}). 

\begin{figure*}
\includegraphics[width=0.55\textwidth]{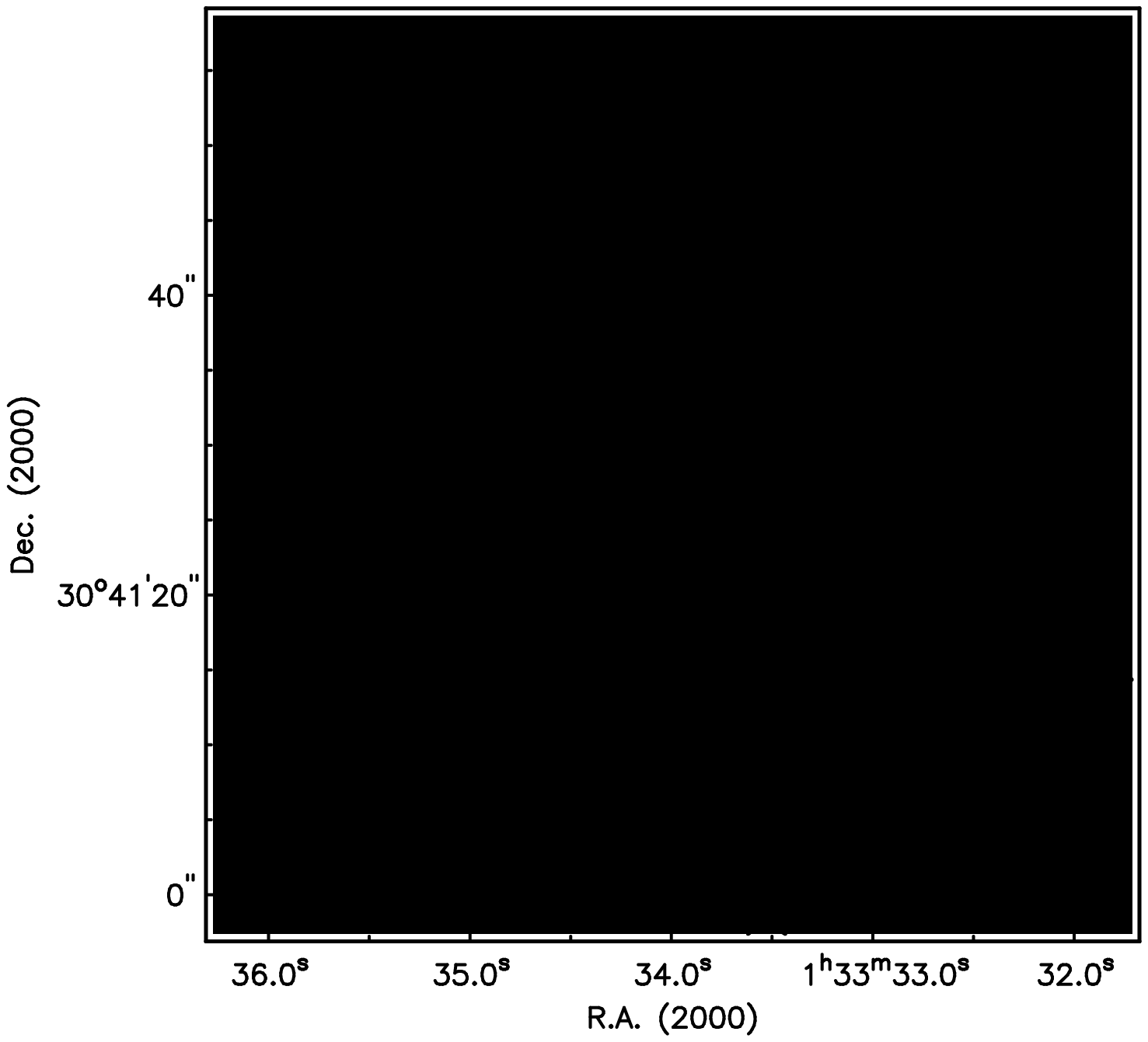}
\includegraphics[width=0.55\textwidth]{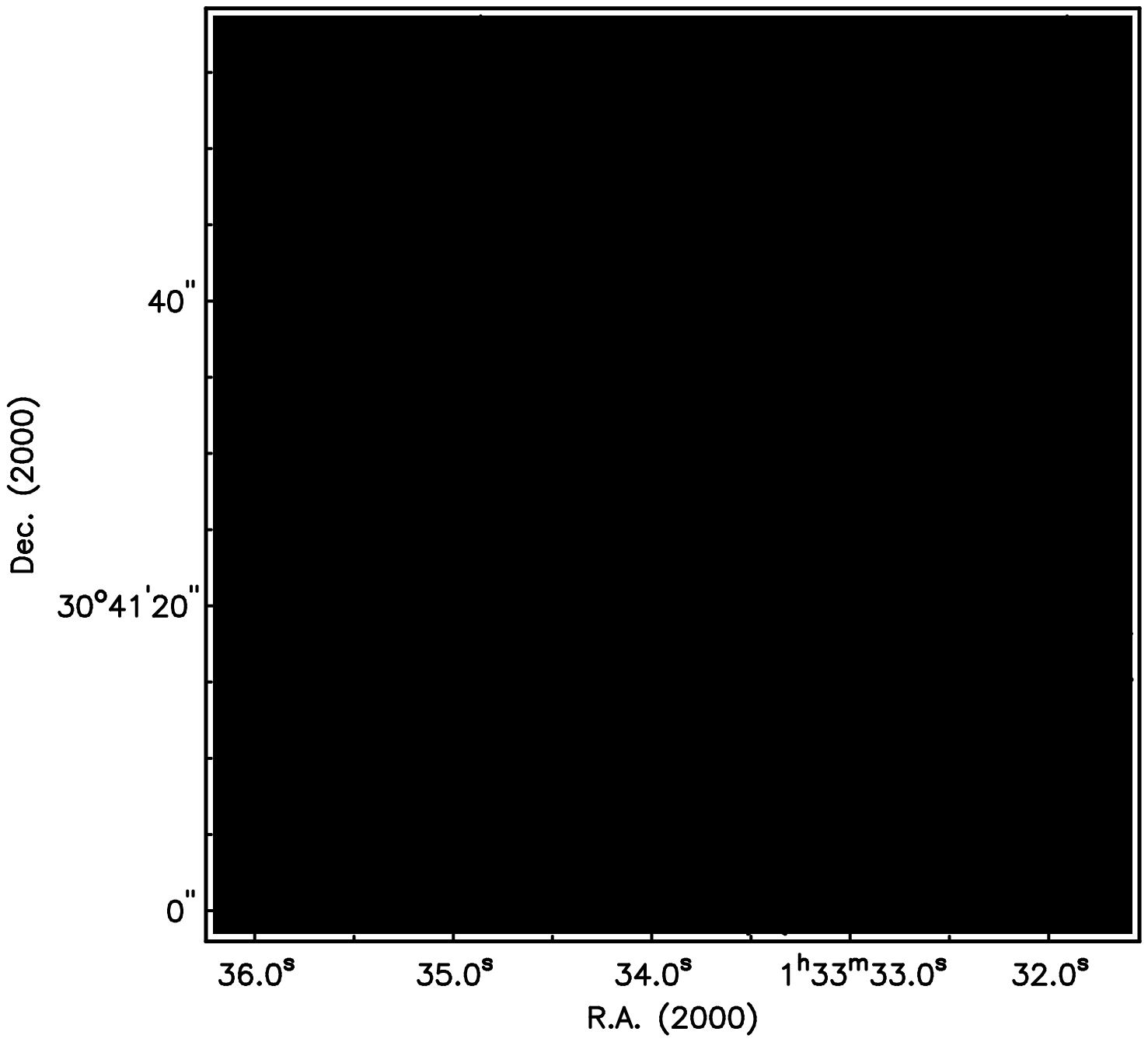}
\includegraphics[width=0.55\textwidth]{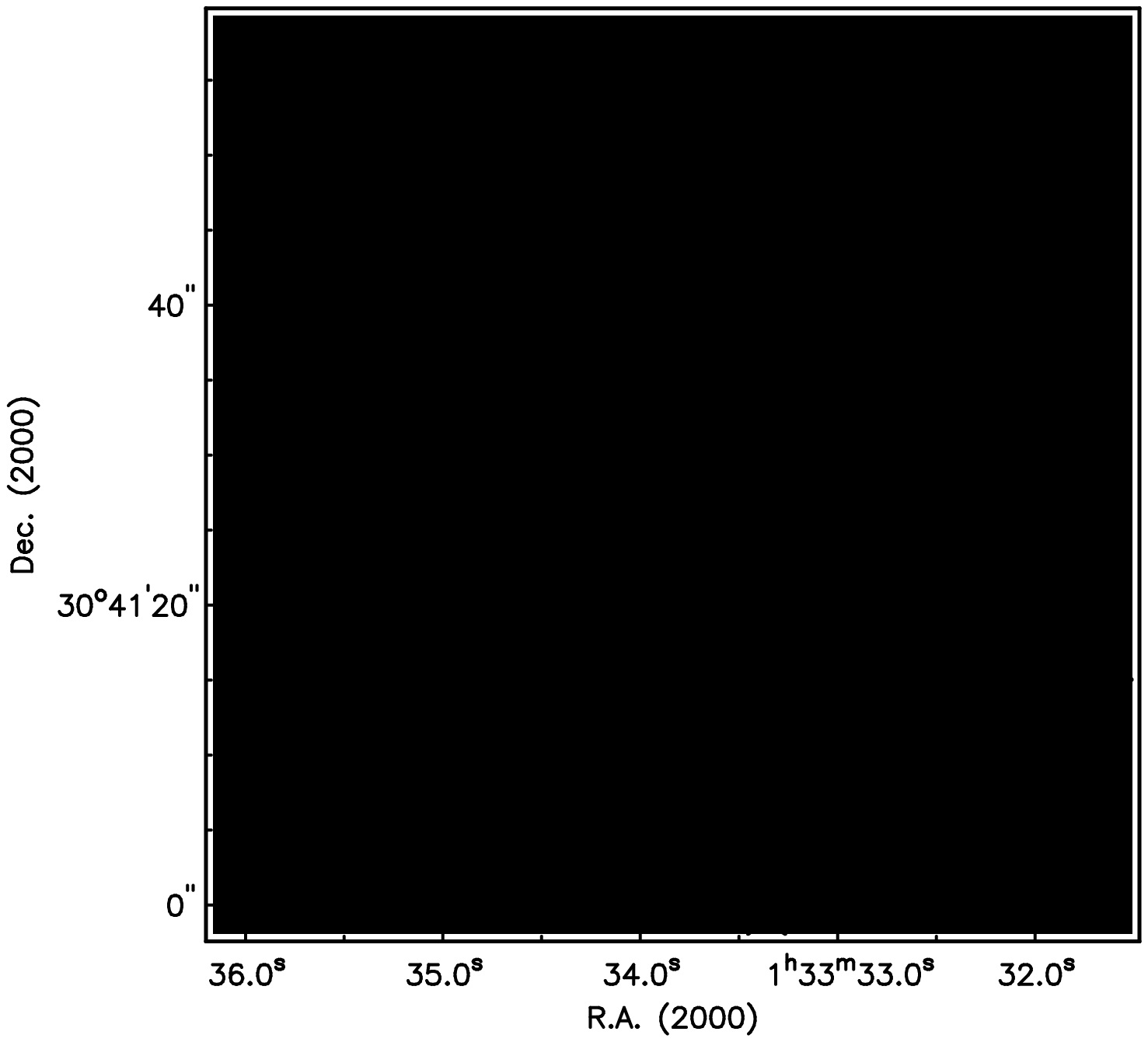}
 \hfill
\includegraphics[width=0.55\textwidth]{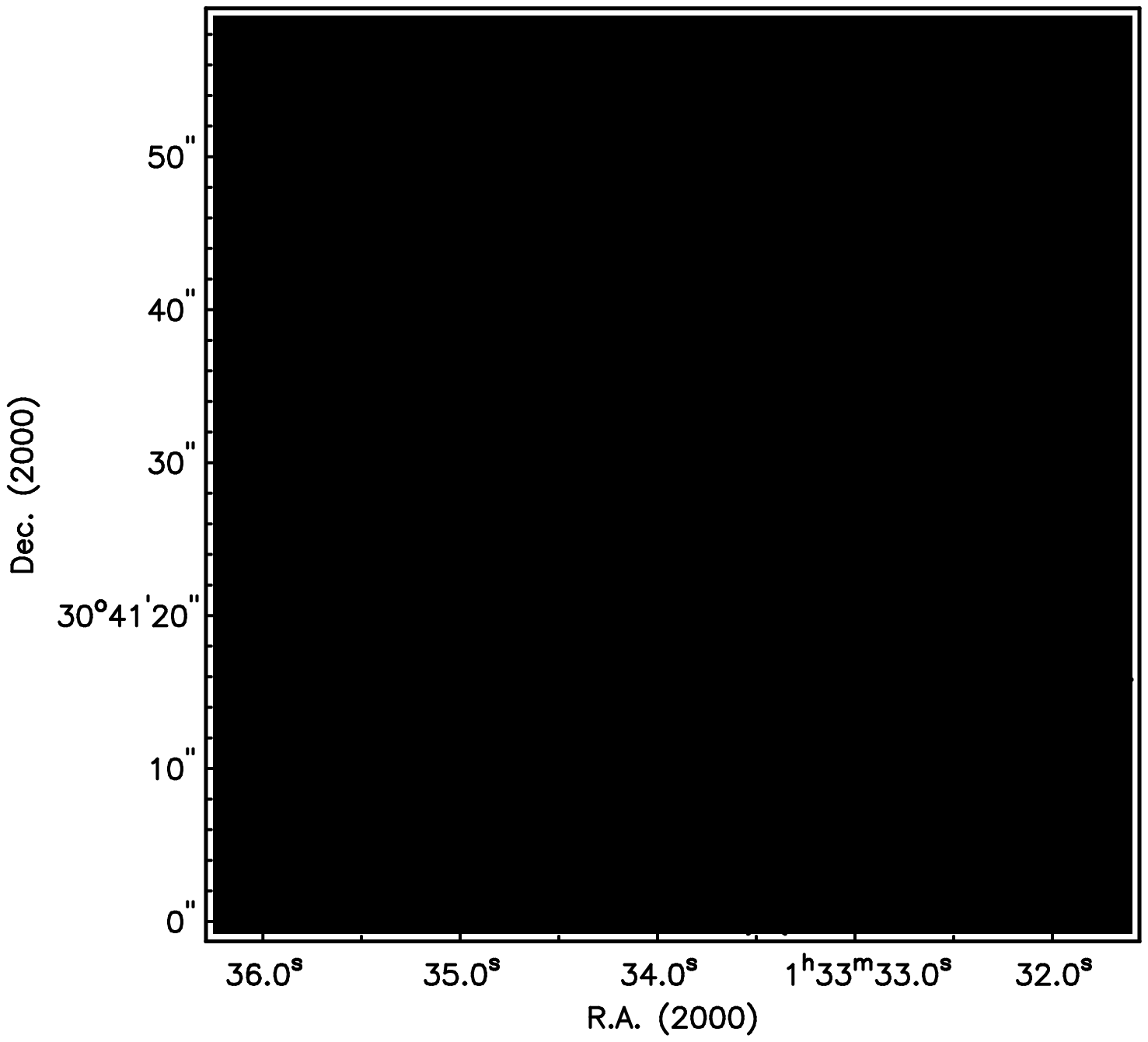}
\caption{Comparison of CO with IR, \ha\  and FUV emission distributions for NGC~595. Contours are the molecular emission, as shown in Wilson \& Scoville (1992) and color images correspond to: top-left: 24\,\mi\ emission, top-right: continuum subtracted \ha\ emission, bottom-left: 8\,\mi\ emission and bottom-right: FUV emission.
In the top-right panel we show the integration zone used to obtain the radial profiles in Figure~\ref{cortes}. The location of the center and the dashed lines ($\pm$45\degree\ from the major axis of the ellipse) were chosen to include the \ha\ emission of the shell structure. The ellipticity was derived assuming an inclination angle of 56\degree\ for M33 (van den Bergh 2000).}
\label{fig7}
\end{figure*}

\begin{figure*}
\plotone{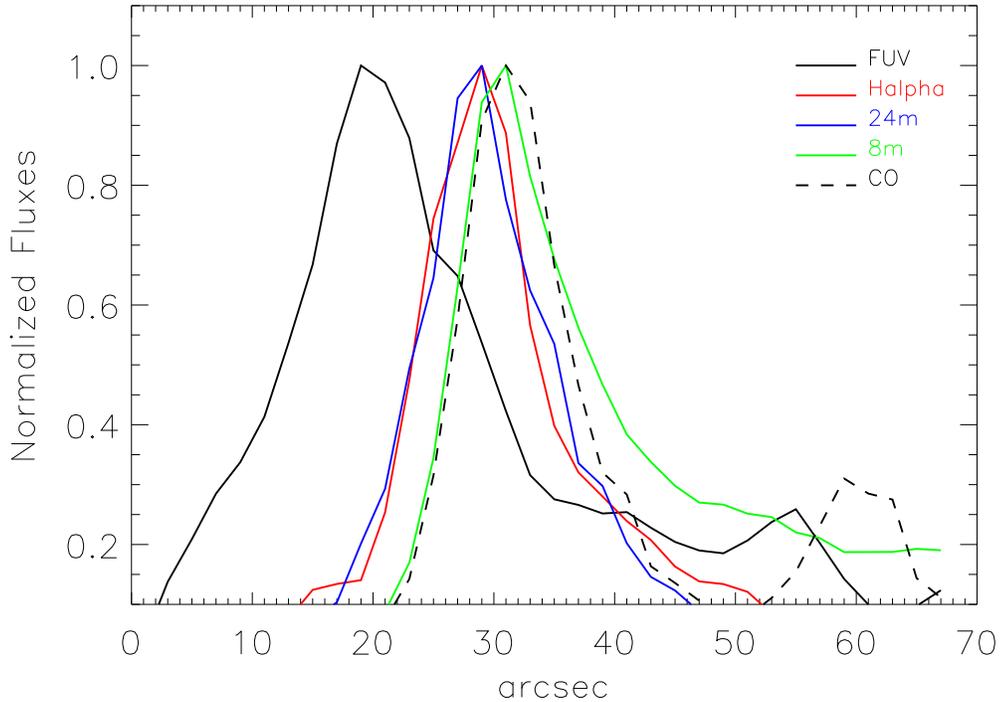}
\caption{Radial profiles of FUV, \ha, 24\,\mi, 8\,\mi\ and CO emission distributions for NGC~595. The integration zone covered by the \ha\ shell structure for this region is depicted in Figure~\ref{fig7}, top-right. The rings are 2\arcsec\ width and the integrated fluxes have been normalized to the maximum value in each ring, the distances are given along the major axis of the ellipse starting from the center position shown as a cross in Figure~\ref{fig7}, top-right. The layered emission distribution is clearly shown here, from the central to the outer radii we find: FUV emission, 24\,\mi\ and \ha\ spatially correlated emissions and 8\,\mi\ and CO emission tracing the outer parts of the \hii\ region.}
\label{cortes}
\end{figure*}

In NGC~595 (Figure~\ref{fig7}), these patterns are even more clearly defined. The figure shows 24\,\mi, \ha, 8\,\mi, and FUV emission with CO contours overlaid. In this region, the CO emission follows perfectly the arched distribution of the 8\,\mi\ emission (bottom left panel in Figure~\ref{fig7}), while the 24\,\mi\ emission distribution shows its maximum slightly shifted from the CO maxima (upper left panel) and is spatially related to the shell-like \ha\ distribution. The FUV emission is clearly shown to be located in the inner part of the \ha-shell structure anticorrelating with the CO emission. The displacement between the different emission lines is better appreciated in Figure~\ref{cortes}. In this figure we show radial profiles extracted over elliptical rings of 2\arcsec\ width covering the complete \ha\ shell structure of NGC~595 (see Figure~\ref{fig7} (top-right) where the integration zone is depicted). The integrated fluxes for each ring have been normalized to the maximum value in the corresponding ring. From the center position to outer radii a layered structure emission is clearly seen: FUV is located at the inner part of the ellipse, then at larger radii there is emission at \ha\ and 
24\,\mi, both following the same distribution, and slightly further out we find the 8\,\mi\ and CO emission distributions. The \ha\ and 24\,\mi\ emission distributions have coincident maxima, as well as the 8\,\mi\ and CO distributions.
The geometry of the shell in NGC~595, defined as a semicircle in \ha, reveals the same layered structure as the one observed in the south-east part of NGC~604, but with the 8\,\mi\  emission much more identified with the molecular cloud. This supports the idea that the 8\,\mi\ emission is indeed tracing the location of the PDR. The emission distribution in the interior of NGC~604 and NGC~595 at the wavelengths we are able to study here, from UV to CO emission, shows that the classical picture of the \hii\ region as seen in the Orion Nebula, with the molecular cloud located at the boundaries of the \hii\ region and delineating the PDR (O'Dell 2001; Hollenbach \& Tielens 1997), also holds for extragalactic \hii\ regions.

%-----------------------------------------------------------------------------%
\subsection{Extinction Maps of NGC~604}
\label{S5.1}
%-----------------------------------------------------------------------------%

\begin{figure*}
\epsscale{0.9}
\includegraphics[width=0.45\textwidth]{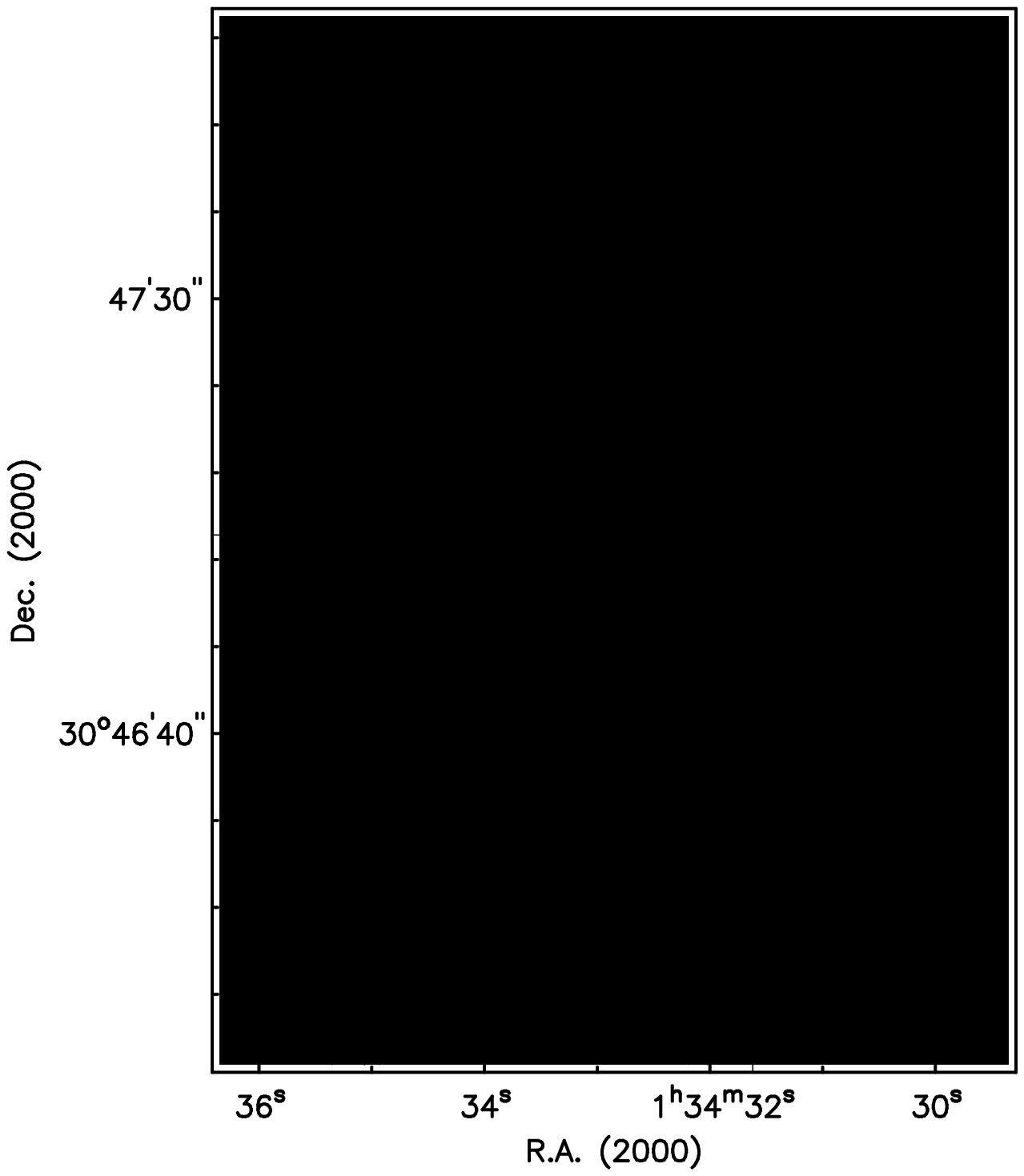}
\includegraphics[width=0.45\textwidth]{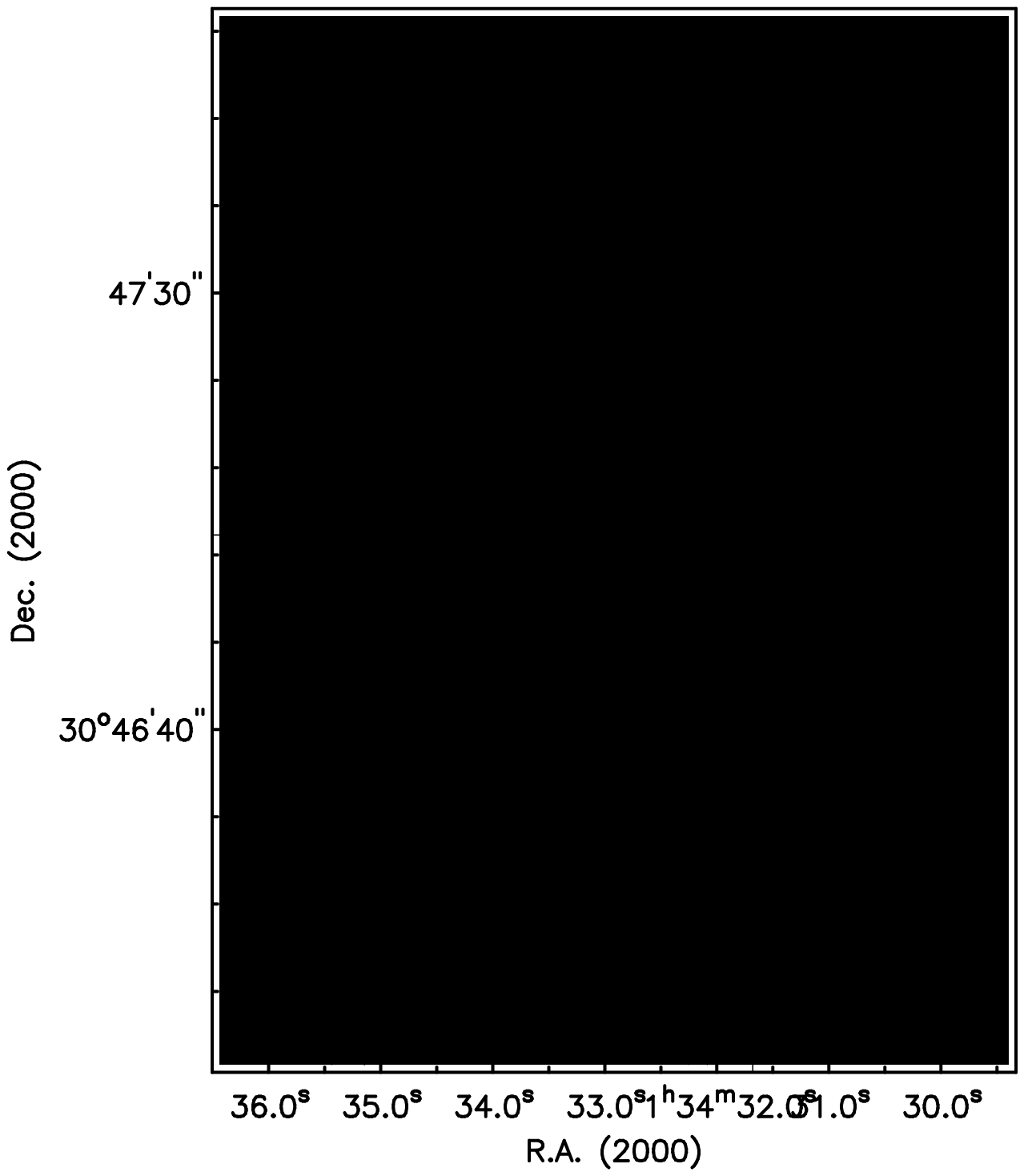}
\caption{Left: Extinction map at 6\arcsec\ resolution obtained from the \ha/\hb\ emission line ratio with CO emission contours overlaid. Right: Extinction map derived from the 24\,\mi\ and \ha\ emission line ratio with CO contours. The \ha\ image has been convolved and regridded to have the same resolution and pixel scale as the 24\,\mi\ image from MIPS, only pixels with S/N$>$15 in the 24\,\mi\ image were considered. The capital letters correspond to the position of radio knots identified in Churchwell \& Goss (1999) (see Figure 7). The black axes inside the plot show the field of view of the CO observations from Wilson \& Scoville (1992). Unfortunately, we are not able to compare both extinction maps with the CO emission for the total \hii\ region, only for the region within the axes.}
\label{fig8}
\end{figure*}

In order to compare the dust emission and extinction we have derived extinction maps for NGC~604 following two methods: one using the Balmer decrement and another based on the ratio between the 24\,\mi\ and \ha\ luminosities. The \ha\ and \hb\ images were smoothed to 6\arcsec\ resolution, and Eq.~(1) was applied with a temperature of 8500~K (Esteban et al. 2002) to derive the extinction map using the Balmer decrement. The result is shown in the left panel of Figure~\ref{fig8}, overlaid with CO emission contours from Wilson \& Scoville (1992) and the identified radio emission knots from Churchwell \& Goss (1999). The Balmer extinction map is similar to the one derived in Bosch et al. (2002) at 2\arcsec\ resolution from \ha\ and \hb\ observations at the 1.0m Jacobus Kapteyn Telescope and to the one derived by 
Ma\' iz-Apell\' aniz et al. (2004) at 4\arcsec\ resolution. The second method applies Eq.~(2) and uses the 24\,\mi\ and \ha\ images of NGC~604. We convolved the \ha\ image to the resolution of the 24\,\mi\ image using the semiempirical PSF for a 75K blackbody\footnote{http://dirty.as.arizona.edu/$\sim$kgordon/mips/conv$_{-}$psfs/conv$_{-}$psfs.html}. Then, the smoothed \ha\ image was regridded to have the same pixel size as the 24\,\mi\ image and Eq.~(2) was applied to derived the extinction map shown in the right panel of  Figure~\ref{fig8}. 

Both extinction maps show localized enhancements that are related to the position of radio knots (marked as capital letters in both figures) and the CO emission. Unfortunately we only have CO observations of the southwest part of NGC~604 (marked as black axes in Figure~\ref{fig8}), thus the comparison between both extinction maps and the CO emission can only be made for that part of the region. Although the range of values for the extinction derived from both methods is quite similar, the extinction distributions within the region are different. 
The comparison of both extinction maps is specially interesting at the position of the main molecular cloud in the region, MC-2 in Wilson and Scoville's notation. 
At this position we find differences derived from both methods. The A$_{24}$(\ha) map shows a maximum that corresponds to the location of MC-2, while in the Balmer extinction map we can see an extinction gradient at this position. A similar phenomenology was observed by Ma\' iz-Apell\' aniz et al. (2004) comparing the Balmer extinction with the extinction derived using radio observations. They found a better correlation between the extinction map derived from radio observations and  the position of the main molecular cloud identified by Engargiola et al. (2003) than between the Balmer extinction map and CO data. Ma\' iz-Apell\' aniz et al. (2004) explain the difference by proposing a scenario in which ''HII gas is located along the surface of the main molecular cloud but this cloud creates a high obscuration ''flap'' that absorbs most of the Balmer photons located behind, letting only radio continuum photons traverse it''. We observe here the same effect, the 24\,\mi\ emission at the position of the MC-2 is produced by the absorption of the ionizing radiation coming from the the stars located behind the flap.  The extinction produced by this flap in the molecular cloud can be studied using radio or 24\,\mi\ emission, as shown in right panel of Figure~\ref{fig8}. The extinction predicted by the Balmer decrement will only account for the extinction suffered by the radiation coming from stars located at the surface of the molecular cloud facing towards us. 

The picture described above agrees well with the results from kinematic studies (Yang et al. 1996). These authors described five \ha\ shells with different sizes and velocities in NGC~604. Their shells identified as 4 and 5 lie next to the position of the MC-2, with part of their surfaces bordering the molecular cloud (see top right panel in Figure~\ref{fig6} in this paper and Figure~2c in Yang et al. 1996). The authors found asymmetric expansion velocities for these shells, as opposed to the symmetric velocities observed in the rest of the shells located away from the molecular cloud. They argue that the expansion of these shells into the molecular clouds would cause the asymetries, and that the shells are ''blisters'' located on the surface of the molecular cloud. If the surface of the molecular cloud has indeed a shell morphology, this 
geometry is consistent with the existence of a high obscuration flap that explains the difference between both extinction maps shown in Figure.~\ref{fig8}.
%-----------------------------------------------------------------------------%
\subsection{Embedded Star Formation in NGC~604?}
\label{S4.2}
%-----------------------------------------------------------------------------%

The difference between the extinction maps explained in the previous section gives clues about the possibility of having an amount of dust along the line of sight
that cannot be detected when only the Balmer extinction is considered. At the location of radio knots A and B Tosaki et al. (2007) found a high CO(J=3-2)/CO(J=1-0) ratio, which reflects conditions of high temperature and density. Using [\sii] $\lambda$6717/$\lambda$6731, Ma\' iz-Apell\' aniz et al. (2004) found high density peaks at the positions of radio knots A and B suggesting the existence of two compact \hii\ regions. These locations could be sites where new embedded star formation is taking place. 

\begin{figure}
\epsscale{0.8}
\plotone{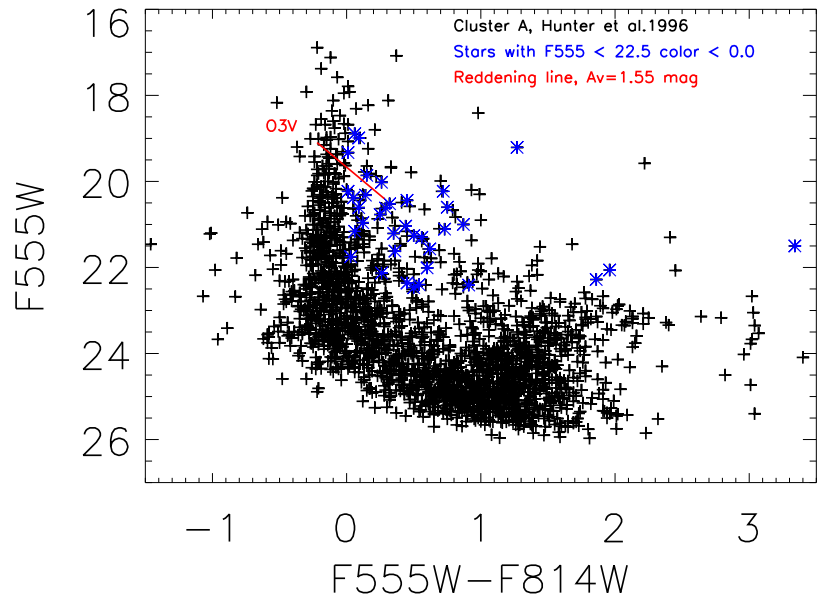}
\caption{Color-magnitude diagram for Cluster A defined in Hunter et al. (1996). The red line shows the extinction correction corresponding to A$\rm _V$=1.5 mag (see text). The stars marked with blue asteriscs are stars located within the molecular cloud MC-2 that have F555W $\leq$ 22.5 and F555W-F814W $\geq$ 0.0. These stars, located to the right of the main sequence, would show an IR-excess and could be classified as young stellar objects.}
\label{fig9}
\end{figure}

\begin{figure}
\epsscale{0.8}
\includegraphics[]{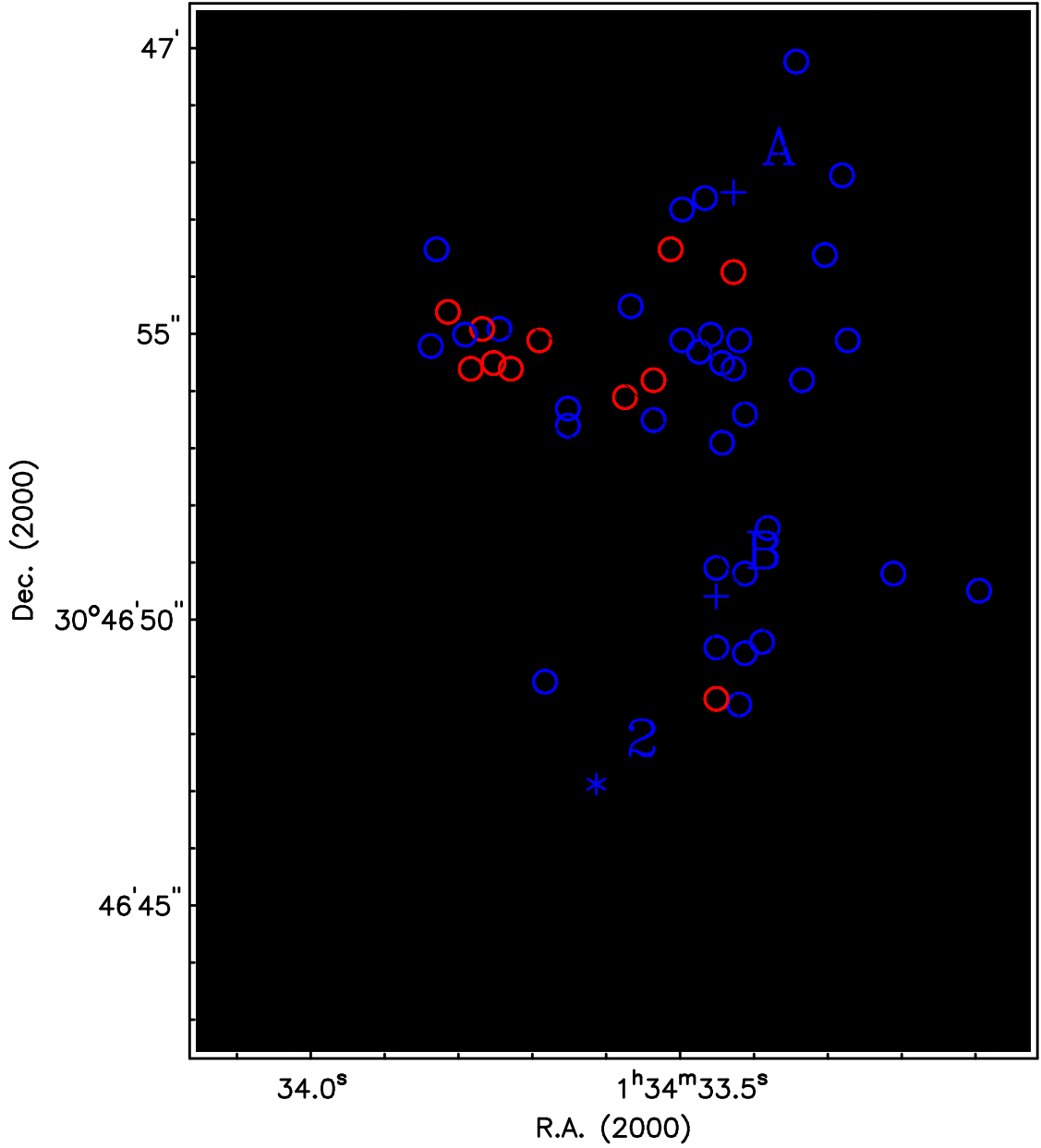}
\caption{F555W image of the zone of NGC~604 where the MC-2 molecular cloud is located. Blue circles show the position of the stars requiring A$\rm _V$ $\leq$ 2.0 to be located on the main sequence in the color-magnitud diagram shown in Figure~\ref{fig9}. Red circles denote stars requiring A$\rm _V>$ 2.0. The contours correspond to A$\rm _V$=1.5, 3.0, 4.5, 6.0, 7.5, 9.0 magnitudes derived from the CO intensity map.}
\label{fig10}
\end{figure}

In order to check this hypothesis we have re-analyzed the stellar photometry presented by Hunter et al. (1996) in a rectangular area shown in Figure~\ref{fig10} that includes the bright radio knots A and B and most of the molecular cloud MC-2. We have used the HST WFPC2 stellar photometry performed by Hunter et al. (1996), which is available through ADC\footnote {Astronomical Data Center: http://adc.astro.umd.edu/adc.html} and created a color-magnitude diagram (CMD) (Figure~\ref{fig9}) of the central star cluster of NGC~604 (Cluster A in Hunter et al.'s notation). The diagram is the same as the one shown in Fig5.a (left panel) in Hunter et al. (1996). No reddening nor distance corrections have been applied. We recognize in the diagram the red giant branch at low F555W luminosity (identified by Hunter et al. 1996 as background stars) and the main sequence for the most luminous stars in the \hii\ region.  We also show the reddening vector corresponding to A$\rm _V$=1.5~mag for an O3(V) star in the main sequence (M$\rm _V=-$5.79, Martins et al. 2002 and $\rm V-I=-$0.322, Bessell et al. 1998), following the extinction curve of Cardelli et al. (1989) with $\rm R_{V}$=3.1 and applying it to the WFPC2 filters (see Holtzman et al. 1995). A$\rm _V=$1.5 mag (E(B$-$V)$\sim$0.5) is a reasonable upper limit of the mean extinction suffered by the stars in NGC~604: it corresponds to the maximum extinction observed in NGC~604 (see Figure~\ref{fig8}) and Churchwell \& Goss (1999) found an overall extinction of  A$\rm _V\sim$0.5 mag for the region. Other studies show lower values for the overall extinction: E(B$-$V)=0.08 (Hunter et al. 1996), E(B$-$V)=0.03 (Pellerin 2006), E(B$-$V)=0.13 (Gonz\'alez-Delgado \& P\'erez 2000), corresponding to A$\rm _V$=0.25, A$\rm _V$=0.09 and A$\rm _V$=0.40, respectively; and here we obtain A$_{\rm Bal}$(\ha)=0.37, which corresponds to A$_{\rm V}$=0.30.

The blue points in Figure~\ref{fig9} represent stars located in a rectangular area including radio knots A and B which are bright (F555W $\leq$ 22.5) and have higher F555W$-$F814W color (F555W$-$F814W$\geq$0.0) than those corresponding to main sequence stars. We have excluded from this selection 8 stars with F555W-F814W errors higher than 0.2~mag, which is the spread seen for the main sequence in this diagram. The stars selected in this way are bright enough not to belong to the red giant branch and they cannot be supergiants (for an O9.5(I) M$\rm _V=-$6.28, Martins et al. (2005), corresponding to an apparent magnitude of m$\rm _V$=18.34). Thus, the stars marked as blue asterisks in Figure~\ref{fig9} could be stars showing a redder excess that are probably forming in dense knots of dust and molecular gas. The location of these stars in NGC~604 is shown in Figure~\ref{fig10} on the HST F555W image from Hunter et al. (1996). 

Another possible reason for the displacement of the stars to the right of the main sequence in the CMD could be the extinction produced by the molecular cloud MC-2, which can cause higher localized extinction than the lower overall extinction in the region. In order to test this possibility, we have derived the extinction required to locate the reddened stars on the main sequence, using the reddening vector shown in the CMD (Figure~10). We find the stars need to be extinction corrected by A$\rm _V\sim$0.1-8.0 mag to be located in the main sequence. 

A crude estimation of the extinction caused by the molecular cloud can be estimated using the conversion of the CO intensity into molecular mass (N(H$_{2}$)/I$_{\rm CO}$=2$\times$10$^{20}$cm$^{-2}$(Kkm s\me)\me, Dickman et al. 1986) and the dust-to-gas mass ratio of  A$\rm _V/N(H)\sim 5.3\times 10^{-22}$mag cm$^{2}$ (Bohlin et al. 1978). The extinction value range derived in this way is  A$\rm _V\sim$1.5-9.0 mag, similar to the extinction range required to locate the stars on the main sequence. In Figure~\ref{fig10} we plot the reddened stars and the extinction contours derived using the molecular emission on the F555W image, blue points are stars which required  A$\rm _{V}<$ 2.0 mag to be located on the main sequence and red points stars requiring  A$\rm _{V}>$2.0. Although the extinction estimates from the CO intensity map are quite crude, they show evidence that the CO molecular cloud could be producing an extra reddening in the \hii\ region that lead us to  confuse the redder stars with stars having an intrinsic IR-excess. 

The reddened stars shown in the CMD could be explained by the existence of a foreground molecular cloud but we cannot rule out the possibility that the radio knots A and B are embedded star forming regions that we cannot detect with our observations. The classical JHK infrared color-color diagram used to detect young stellar objects (Lada \& Adams 1992) could be useful here to search for evidence of embedded star formation. The principal problem in using JHK photometry for NGC~604 is the detection limit; young stellar objects such as T Tauri stars, Herbig AeBe stars and Class I sources can be very difficult to see at the distance of M33, because their apparent magnitudes range between m$\rm _J\sim$26 for T Tauri stars to m$\rm _J\sim$22 for Class I sources (see Figure~3 in Brandner et al. 2001). 

Although we are not able to confirm the existence of embedded star formation in NGC~604 using the observations presented in this paper, we show here results that reinforce this hypothesis. First, the locations of the reddened stars coincide with high CO and 8\,\mi\ emission, second, we need to assume that there is ionized gas completely hidden by dust to explain the differences between the extinctions derived from the Balmer decrement and the extinction derived using 24\,\mi\ (Figure~\ref{fig8}). Third, the high CO(J=3-2)/CO(J=1-0) ratio found by Tosaki et al. (2007) at the position of the radio knots A and B. All these results suggest that there should be 
embedded star formation in the surroundings of the molecular cloud MC-2. 

%-----------------------------------------------------------------------------%
\section{Summary and Discussion}
\label{S6}

%-----------------------------------------------------------------------------%

We present observations to study the dust and gas emission distribution within a set of the most luminous \hii\ 
regions in M33. The linear resolution of the observations allows us to make a 
comparison of the 24\,\mi, 8\,\mi, \ha\ and UV emissions in the interior of the \hii\ region sample and quantify the star formation for each individual \hii\ region using different wavelengths. We find the following results: 

\begin{itemize}
\item[1.] The general picture assumed in previous extragalactic studies, which suggests a correspondence between the 24\,\mi\ emission and the nebular \ha\ emission line, is found here at the small local scales where the stars form 
in the \hii\ regions. The 24\,\mi\ emission structure suggests that the emitted dust is mixed with the ionized gas in the interior of the \hii\ regions, while the 8\,\mi\ emission is associated with the \hii\ region boundaries and neutral material. The UV emission is in general not associated with 24\,\mi\ and 8\,\mi\ emissions, and it is clearly surrounded by the emission of the ionized gas at \ha. 

\item[2.] We quantify the SFs for our sample of \hii\ regions using a set of available wavelengths and assuming an instantaneous burst of SF. The SFs derived from the 24\,\mi\ emission are up to a factor of 10 lower than that derived from the extinction--corrected \ha\ luminosity, while the SFs derived using a linear combination of 24\,\mi\ and observed \ha\ emission give better estimations of the total SFs in the regions than the estimations from the 24\,\mi\ or 8\,\mi\ emissions by themselves.  

\item[3.]  The 8\,\mi\ emission fails to reproduce the SFs values derived using the extinction--corrected \ha\ luminosity, which is not surprising since 8\,\mi\ is more associated with the boundaries of the \hii\ regions and shows larger systematic uncertainties and a larger scatter when it is calibrated as a SFR tracer. 

\item[4.]  We show that the observed UV and \ha\ luminosities are consistent with young stellar populations ($\tau\sim$3-4\,Myr) and that the SFs predicted from extinction--corrected FUV fluxes are similar to the values derived from \ha\ when the instantaneous SF approximation is taken into account. 

\item[5.]  We have derived extinction maps for NGC~604, the most luminous \hii\ region in our sample. We see that although global extinction is modest (A$_{V}$=0.30), we do find localized regions within NGC~604 with much higher extinction. The extinction values estimated from the 24\,\mi\ and \ha\ luminosity ratio are higher than those derived using the Balmer decrement at the location of MC-2. The underestimated extinction values using the Balmer method are probably related  to geometrical effects at the surface of the main molecular cloud in NGC~604. In spite of these differences, the extinction variations seem to have little effect on global extinction derivations and thus on the SF measurements presented here. 

\item[6.]  The SF derived from the 24\,\mi\ emission better traces the extincted SF in the \hii\ regions. This is supported by the good spatial correlation between the absorbed \ha\ luminosity of NGC~604, derived using a Balmer extinction map of the region, and the emission at 24\,\mi. This is expected if the dust emitting at 24\,\mi\ reprocesses the light coming from the ionizing stars. 

\end{itemize}

The results shown in this paper have important implications for the study of the star formation in galaxies. Our sample of \hii\ regions are typical of star forming regions in normal galaxies, therefore we expect the conclusions derived here at local scales to hold when studying statistically the SF in big galaxy samples. We see that: a) since dust reprocesses only a fraction of the UV stellar emission, the SF derived using IR emission (24\,\mi, 8\,\mi\  and IR) will underestimate the SF. b) Although at small scales there are differences in the extinction derived from the Balmer decrement and from the 24\,\mi-to-\ha\ luminosity ratio for NGC~604, the integrated values for the extinction derived from both methods agree within the uncertainties (A$_{\rm Bal}$(\ha)=0.37$\pm$0.16, A$_{\rm 24}$(\ha)=0.40$\pm$0.07). This supports the hypothesis of using 24\,\mi\ emission to correct the \ha\ luminosity for extinction. The local differences in extinction can be used to find places where there can be a higher dust content than the one derived only from the extinction of optical emission lines. c) We do not see a similar correspondence between 24\,\mi\ and UV emission, which shows the inability of using the 24\,\mi\ emission to correct UV fluxes for extinction. d) We also note here that the SFR calibrations derived assuming continuous SF approximation, generally applied to galaxies, are not applicable in \hii\ regions. We show that an instantaneous burst of SF is a better approximation for our objects and we present the corresponding SFs calibrations for a burst of age 4~Myr. 

The detailed study presented here using high-resolution data allows us to disentangle the spatial correspondence between the 24\,\mi\ and 8\,\mi\ emissions, which have been recently proposed as tracers of the SFR, and the location where the stars form. This study shows that the assumptions on which the statistical studies rely are valid and that the conclusions inferred from them also hold at small galactic scales. In spite of this, absolute SFR calibrations generally used for galaxies have to be treated with extreme care when applying to local star forming regions such as individual \hii\ regions. 

It would be useful to extend the approach presented here to a larger sample 
of \hii\ regions in M33 and the Magellanic Clouds. We would also like to study the dust emission at longer wavelengths using the same method as presented here, especially to compare the 24\,\mi\ emission associated with the ionizing stars with cooler dust emitting at 70\,\mi\ and 160\,\mi, but the resolutions of MIPS is insufficient to carry out this project in our sample of \hii\ regions. Our limited comparison with CO observations illustrates the power of understanding the nebular geometry of the \hii\ region (see also Ercolano et al. 2007) and the general structure of the interior of the \hii\ regions, but more data (e.g. Herschel and HI observations) would be helpful to extend the conclusions reached here.

%-----------------------------------------------------------------------------%
\acknowledgements
%-----------------------------------------------------------------------------%
We would like to thank the anonymous referee for the many constructive comments that have helped improve the manuscript. We would like to thank Christine Wilson for kindly providing us with the CO data. We also thank Caina Cao, Ute Lisenfeld and Almudena Zurita for useful discussions. This work has been carried out under the postdoctoral contract granted by the Programa de Ayudas de Movilidad of the Spanish MICINN, reference EX2007-0985. This research was supported by a Marie Curie Intra European Fellowship within the 7$^{\rm th}$ European Community Framework Programme. Some of the data presented in this paper were obtained from the Multimission Archive at the Space Telescope Science Institute (MAST). STScI is operated by the Association of Universities for Research in Astronomy, Inc., under NASA contract NAS5-26555. Support for MAST for non-HST data is provided by the NASA Office of Space Science via grant NAG5-7584 and by other grants and contracts. This work is based in part on observations made with the Spitzer Space Telescope, which is operated by the Jet Propulsion Laboratory, California Institute of Technology under a contract with NASA. This research draws upon data provided by {\it The Resolved Stellar Content of Local Group Galaxies Currently Forming Stars} PI: Dr. Philip Massey, as distributed by the NOAO Science Archive. NOAO is operated by the Association of Universities for Research in Astronomy (AURA), Inc. under a cooperative agreement with the National Science Foundation.

%-----------------------------------------------------------------------------%

%%%%%%%%%%%%%%FIGURES%%%%%%%%%%%%%%%%%%
\clearpage

\clearpage

\clearpage

\clearpage

\clearpage

\clearpage

\clearpage

\clearpage
\clearpage

\clearpage

\clearpage

%%%%%%%%%%TABLES%%%%%%%%%%%%%%%%
\clearpage
\begin{landscape}
\begin{deluxetable}{lccccccccc} 
  \tabletypesize{\scriptsize} 
%  \rotate
  \tablecaption{Aperture Photometry} 
  \tablewidth{0pt} 
  \tablehead{
  \colhead{Region} &
  \colhead{R.A.} &
  \colhead{Decl.} &
   \colhead{Radius} &
  \colhead{log L$_{obs}$(FUV)\tablenotemark{a}} &
  \colhead{log L$_{obs}$(NUV)\tablenotemark{a}} &
  \colhead{log L(8\,\mi)\tablenotemark{a}} & 
  \colhead{log L(24\,\mi)\tablenotemark{a}}  &
  \colhead{log L$_{obs}$(\ha)\tablenotemark{a}}  &
  \colhead{log L$_{corr}$(\ha)\tablenotemark{a}}   \\
  \colhead{} &
  \colhead{J2000.0} &
  \colhead{J2000.0} &
   \colhead{(\arcsec)} &
  \colhead{(\ergs)} &
  \colhead{(\ergs)} &
  \colhead{(\ergs)} &
  \colhead{(\ergs)}   & 
  \colhead{(\ergs)} & 
  \colhead{(\ergs)}   \\
  }
   \startdata 
  NGC~588 & 01\,\ 32\,\ 45.5 & 30\,\ 38\,\ 56  & 35 & 39.895$\pm$0.004 & 39.795$\pm$0.004 &  38.85$\pm$0.06 &  38.98$\pm$0.04 &  38.70$\pm$0.02  &   38.92$\pm$0.09  \\
  NGC~592 & 01\,\ 33\,\ 12.1\,\ &  30\,\ 38\,\ 48 & 35 & 40.005$\pm$0.004 &39.942$\pm$0.004  &   39.22$\pm$0.08 &  39.19$\pm$0.09 &   38.41$\pm$0.05 &   38.55$\pm$0.12 \\
  IC131       & 01\,\ 33\,\ 16.0 &  30\,\ 45\,\ 10 & 44  & 39.638$\pm$0.016 &39.490$\pm$0.022 & 39.17$\pm$0.14 & 38.84$\pm$0.07 & 38.45$\pm$0.05 &       38.52$\pm$0.05  \\
  IC131-West & 01\,\ 33\,\ 11.4 &  30\,\ 45\,\ 15 & 22 & 39.604$\pm$0.004 & 39.492$\pm$0.004&  39.20$\pm$0.03 &   39.30$\pm$0.02 &  38.34$\pm$0.01 &   38.41$\pm$0.01  \\
  NGC~595 & 01\,\ 33\,\ 33.9 & 30\,\ 41\,\ 40  & 44 & 40.148$\pm$0.005 & 40.028$\pm$0.08 &  39.99$\pm$0.09 &   40.18$\pm$0.04 &  39.10$\pm$0.02 &   39.21$\pm$0.02 \\
  NGC~604 & 01\,\ 34\,\ 33.0 &  30\,\ 47\,\ 05 & 50  & 40.808$\pm$0.001 & 40.668$\pm$0.013 & 40.47$\pm$0.05 &   40.64$\pm$0.01 &  39.49$\pm$0.01 &   39.63$\pm$0.07 \\
  \enddata
   \tablenotetext{a}{The errors correspond to the uncertainties in the flux estimations due to the noise of the image. For the extinction 
   corrected \ha\ luminosity the error is the combination of the flux uncertainty and the error in the extinction (column~2 in Table~2).} 
\label{t3}
\end{deluxetable} 
\clearpage
\end{landscape}

\clearpage
%\begin{landscape}
\begin{deluxetable}{lccccc} 
  \tabletypesize{\scriptsize} 
%  \rotate
  \tablecaption{Extinctions} 
  \tablewidth{0pt} 
  \tablehead{
  \colhead{Region} & 
   \colhead{A$_{\scriptsize Bal}$(\ha)\tablenotemark{a}} &
   \colhead{A$_{\scriptsize 24}$(\ha)\tablenotemark{b}} &
   \colhead{A$_{\scriptsize rad}$(\ha)\tablenotemark{c}} & 
   \colhead{A(FUV)\tablenotemark{d}} &  
   \colhead{References\tablenotemark{e}}  \\
  \colhead{} & 
  \colhead{mag} &
  \colhead{mag} &
  \colhead{mag} &
  \colhead{mag} &
 \colhead{}    \\
  }
   \startdata
  NGC~588 &   0.56$\pm$0.22   & 0.06$\pm$0.01   &    ---                    &  2.08$\pm$0.74   &1,2,3\\
  NGC~592 &   0.35$\pm$0.26   & 0.18$\pm$0.05   &  0.72$\pm$0.40  &  2.55$\pm$0.74   &2,3\\
  IC131        &   0.17    		& 0.08$\pm$0.02   &  0.26$\pm$0.25 &  1.46$\pm$0.94  &4 \\
  IC131-West & 0.17    		& 0.27$\pm$0.05  &  0.52$\pm$0.23 &   1.90$\pm$0.74   &4 \\
  NGC~595 &    0.27$\pm$0.01  & 0.34$\pm$0.06  &  0.51$\pm$0.16 &  1.80$\pm$0.75  & 5 \\
  NGC~604 &    0.37$\pm$0.16   & 0.40$\pm$0.07  &  0.61$\pm$0.08 &  1.55$\pm$0.73   & This paper \\
  \enddata
  \tablenotetext{a}{A$_{\scriptsize Bal}$(\ha) is the value adopted here to correct the observed \ha\ luminosities. For each \hii\ region, the extinction 
  is the mean value of the extinctions reported in the references listed in the last column. The errors in the magnitudes are the standard deviations 
  of the different values given in the literature. For NGC~595 we take the value obtained from the Balmer decrement using PMAS observations (Rela\~no et al. 2009) and for NGC~604 the value derived in this paper. The extinctions reported here are total extinctions, assuming a foreground reddening of E(B-V)=0.07 (van den Bergh 2000) and a Cardelli et al. (1989) extinction law, we derive a value of  A(\ha)=0.17~mag for the Galactic extinction.} 
   \tablenotetext{b}{The extinction uncertainties referred to uncertainties in the measured fluxes (see Table~1 and text).}
  \tablenotetext{c}{The extinction errors are a combination of the errors in the measured \ha\ and radio fluxes and an uncertainty in the estimated aperture radio of 20\%, except for IC131 where an uncertainty of 10\% is considered due to the extended \ha\ shell structure not related to radio emission. The uncertainties for the radio fluxes are those given in Gordon et al. (1999). An additional uncertainty in 
  A$_{\scriptsize rad}$(\ha) is related to changes in the \hii\ region temperature;  A$_{\scriptsize rad}$(\ha) will decrease by 0.1~mag when the 
  temperature increases by $\sim$1500~K.}
\tablenotetext{d} {The FUV extinctions are derived using Starburst99 models (Leitherer et al. 1999) (see  \S 4.3). The errors quoted here are the combination of photometric errors and the uncertainties in the FUV and NUV zero-point calibration,  $\pm$0.05~m$_{AB}$ and $\pm$0.03~m$_{AB}$ for FUV and NUV, respectively (Morrissey et al. 2007). The values shown here are 
  total extinctions. Assuming Cardelli et al. (1989) extinction law with R$_{V}$=3.1, we estimate a contribution for the Galactic foreground of A(FUV)=0.58~mag.}
\tablenotetext{e}{References for A$_{\scriptsize Bal}$(\ha): 1.- Melnick (1979), 2.- Viallefond \& Goss (1986), 3.- Melnick et al. (1987), 4.- V\'ilchez et al. (1988), 5.- Rela\~no et al. (2009)} 
\label{t3}
\end{deluxetable} 
\clearpage
%\end{landscape}

\clearpage
\begin{landscape}
\begin{deluxetable}{lcccccccc} 
%\hspace{-1cm}
  \tabletypesize{\scriptsize} 
   %\rotate
  \tablecaption{Star Formation for the selected \hii\ regions in M33} 
  \tablewidth{0pt} 
  \tablehead{
  \colhead{Region} &
  \colhead{SF(8\,\mi)\tablenotemark{a}} & 
  \colhead{SF(24\,\mi)\tablenotemark{b}} & 
  \colhead{SF(\ha$^{\rm corr}$)\tablenotemark{c}}  &
  \colhead{SF(com)\tablenotemark{d}} & 
  \colhead{SF(FUV$^{\rm corr}$)\tablenotemark{e}}  &
  \colhead{SF(24\,\mi)/SF(\ha$^{\rm corr}$)} & 
  \colhead{SF(com)/SF(\ha$^{\rm corr}$)}  & 
  \colhead{SF(\ha$^{\rm corr}$)/SF(FUV$^{\rm corr}$)}    \\
  \colhead{} &
  \colhead{$10^{4}$\msun} & 
  \colhead{$10^{4}$\msun} & 
  \colhead{$10^{4}$\msun}  &
  \colhead{$10^{4}$\msun} & 
  \colhead{$10^{4}$\msun}  &
  \colhead{}  & 
  \colhead{}  &
  \colhead{} \\
  }
   \startdata
NGC~588     &  0.07$\pm$0.01 &   0.97$\pm$0.11 &   10.85$\pm$2.24 &  6.86$\pm$0.26 & 4.92$\pm$3.36 & 0.09$\pm$0.02 & 0.63$\pm$0.13 & 2.20$\pm$1.58   \\
NGC~592     &  0.19$\pm$0.04 &   1.49$\pm$0.32 &   4.63$\pm$1.24   &  3.98$\pm$0.44 & 9.82$\pm$6.71 & 0.32$\pm$0.11 & 0.86$\pm$0.25 & 0.47$\pm$0.36   \\
IC131           &   0.16$\pm$0.05 &   0.73$\pm$0.11 &   4.29$\pm$0.50  &  3.94$\pm$0.43  &  1.54$\pm$1.33 & 0.17$\pm$0.03 & 0.92$\pm$0.15 &  2.79$\pm$2.43   \\
IC131-West  &  0.18$\pm$0.01 &   1.86$\pm$0.08 &   3.31$\pm$ 0.11  & 3.63$\pm$0.18 &  2.15$\pm$1.46 & 0.56$\pm$0.03 & 1.10$\pm$0.07 &  1.53$\pm$1.05 \\
NGC~595     &  1.78$\pm$0.41 & 11.16$\pm$0.94 &   21.00$\pm$1.06 &  22.42$\pm$1.51 & 6.86$\pm$4.72 & 0.53$\pm$0.05 & 1.07$\pm$0.09 &  3.06$\pm$2.11 \\
NGC~604     &  7.16$\pm$0.90 & 28.43$\pm$0.82 &  55.55$\pm$8.34 & 56.87$\pm$3.60  & 24.78$\pm$16.59 & 0.51$\pm$0.08 & 1.02$\pm$0.17& 2.24$\pm$1.54  \\  
     \enddata
 \tablenotetext{a}{Using the 8\,\mi-\ha\ relation fitted by Calzetti et al. (2005) and then applying the calibration for an instantaneous burst of SF: 
 SF(\ha)(\msun)$\rm =1.29\times10^{-34}L(\ha)(\ergs)$, see Eq. (3). All the SF are given in $10^{4}$\msun. The errors given here account for the photometric errors and 
 the errors for the extinctions quoted in Table~2.} 
 \tablenotetext{b}{Using the empirically relation between $\rm L_{24}(\ergs)$ and $\rm L_{corr}(\ha)(\ergs)$ given by Calzetti et al. (2007)} 
 \tablenotetext{c}{Using SF calibration for an instantaneous burst of SF (see Eq. (3)).} 
 \tablenotetext{d}{Assuming $\rm {L_{corr}(\ha)=L_{obs}(\ha)+0.031\times L_{24}}$ 
 			and then applying Eq. (3).} 
\tablenotetext{e}{Applying the SF calibration for the FUV emission for an instantaneous burst of SF at $\tau$=4~Myr
derived in this paper; SF(FUV)(\msun)$\rm =1.64\times10^{-21}L(FUV)(\ergs\ Hz\me)$.} 
\label{t2}
\end{deluxetable} 
\clearpage
\end{landscape}
\end{document}